\documentclass[letterpaper,twocolumn,10pt]{article}
\usepackage{usenix-2020-09}

\usepackage{url}
\usepackage{fancyhdr}
\usepackage{graphicx}
\usepackage{listings}
\usepackage{color}
\usepackage{xcolor}
\usepackage{tikz}
\usepackage{booktabs}
\usepackage{float}
\usepackage{xspace}
\usepackage{adjustbox}
\usepackage{mathtools}
\usepackage{pifont}
\usepackage[caption=false,font=normalsize,labelfont=sf,textfont=sf]{subfig}
\usepackage{multirow}
\usepackage{mdframed}
\usepackage{hyperref}
\usepackage[
all=normal%
,floats=tight%
,paragraphs=tight%
,wordspacing=tight
]{savetrees}

\usepackage[backend=biber,isbn=false,url=false,doi=false,style=numeric,sorting=none,maxbibnames=99,giveninits=true]{biblatex}

\microtypecontext{spacing=nonfrench}

\usetikzlibrary{positioning,calc,arrows,decorations,shapes,decorations.pathreplacing,chains,fit,arrows.meta,intersections,positioning,automata,external,matrix,tikzmark}

\usepackage[firstpage]{draftwatermark}
\SetWatermarkText{\hspace*{6in}\raisebox{23.5cm}{\includegraphics{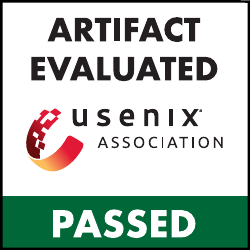}}}
\SetWatermarkAngle{0}

\input{definitions}

\title{Frontal Attack: Leaking Control-Flow in SGX via the CPU Frontend}

\author{Ivan Puddu, Moritz Schneider, Miro Haller, Srdjan \v{C}apkun\\
    Department of Computer Science\\
    ETH Zurich}

\begin{document}

\maketitle

\begin{abstract}
We introduce a new timing side-channel attack on Intel CPU processors. Our \emph{\attackname} attack exploits timing differences that arise from how the CPU frontend fetches and processes instructions while being interrupted. In particular, we observe that in modern Intel CPUs, some instructions' execution times will depend on which operations precede and succeed them, and on \emph{their virtual addresses}. Unlike previous attacks that could only profile branches if they contained different code or had known branch targets, the \attackname attack allows the adversary to distinguish between instruction-wise identical branches. As the attack requires OS capabilities to set the interrupts, we use it to exploit SGX enclaves. Our attack further demonstrates that secret-dependent branches should not be used even alongside defenses to current controlled-channel attacks.
We show that the adversary can use the \attackname attack to extract a secret from an SGX enclave if that secret was used as a branching condition for two instruction-wise identical branches. We successfully tested the attack on all the available Intel CPUs with SGX (until 10th gen) and used it to leak information from two commonly used cryptographic libraries.
\end{abstract}

\section{Introduction}
Today's computing world runs in the cloud. Massive data centers maintained by cloud providers are the infrastructure upon which companies and most of the internet are increasingly relying~\cite{cloudrise}. For many use cases, renting computing resources is cost-effective and convenient. Resources can dynamically scale up when demand is high, while not having to maintain them.
Security-wise, on the other hand, cloud computing is a much harder sell. Offloading computation and data to a third party raises questions about confidentiality and integrity. Not only can a remote attacker rent the same server and be co-located with the victim, but the provider itself could be malicious. In such a scenario, hypervisors and operating systems (OS), which usually provide isolation, can be easily compromised and thus offer little to no assurance in terms of security. 

This setting has been a driving force in recent efforts to develop trusted execution environments (TEEs). While there are many TEE proposals~\cite{sgxexplained,trustzone,keystone,sancus,sanctum,komodo,amdsev}, they are unified in their goal: providing an integrity and confidentiality oasis in an environment ruled by malicious operating systems and hypervisors. The fundamentals for this oasis are rooted in the lowest level of the computing stack: the CPU. When application security is provided through CPU primitives, the layers above need not be trusted. %
Intel SGX~\cite{sgxexplained} is the most widely deployed among all the TEE proposals, being available in almost every modern consumer CPU Intel manufactures. It protects applications by running them in \emph{enclaves}. SGX authenticates and encrypts enclaves' memory accesses that cross the CPU boundary and blocks any other software in the system, including OS and hypervisor, from accessing enclaves' code and data.
Nevertheless, as protected as they might be, enclaves do not execute in isolation. Enclaves share resources with other applications running in the same system, particularly memory and CPU time. By design, SGX leaves the (untrusted) OS in charge of managing these resources.

However, whenever shared resources are involved, so are side-channels. Researchers were quick to point out this shortcoming of SGX~\cite{sgxexplained,kari2017softwaregrandexposure,l1sgxcacheattack,cacheattacksgx17,ccpagefault}, casting doubt into enclaves' ability to provide confidentiality, one of the core TEE goals. Intel acknowledged the problem but shifted the burden of protecting against side-channels to enclave developers~\cite{sgxsidechannelstance}.
Curbing side-channels is not trivial, and in the case of SGX, it is particularly challenging due to the role the OS plays. To manage the system resources, the OS is responsible for the enclave scheduling, memory paging, and interrupt and exception handling, to name a few. These OS capabilities, which the attacker controls, decrease the noise of traditional side-channel attacks~\cite{kari2017softwaregrandexposure,lee2017inferring} and enable new types of side-channels, called controlled-channel attacks~\cite{ccpagefault}.

The first controlled-channel attacks allowed the adversary to observe enclave accesses at page granularity (4 KiB) without any noise by merely abusing memory paging. Revoking permissions to the enclave's pages leads to page-faults, which in turn give the OS attacker a trace of every page the enclave accesses. %
Initial defenses that worked on the assumption that the attacker would need to trigger page-faults~\cite{tsgxshih17}, just prompted the emergence of stealthier attacks that observe page metadata set by the CPU~\cite{leakycaulderon17,accesscontrolledchannel}. %
In response to these attacks, Intel officially recommends SGX developers place sensitive data and code within a page~\cite{sgxsidechanneldev}.
Controlled channels, however, do not stop at the page boundary. OS capabilities can be used to enhance cache attacks~\cite{kari2017softwaregrandexposure,l1sgxcacheattack,cacheattacksgx17} and extract enough information from the branch prediction unit (BPU) to give the attacker a branch granularity view of the victim~\cite{lee2017inferring,evtyushkin2018branchscope,bluethunder}. As this undermines defenses against paging-based controlled channels, further defenses leveraged the coarse timing resolution of the attacker and the inability of BPU attacks to leak the target of unconditional branches~\cite{lee2017inferring}. %
Nemesis~\cite{van2018nemesis} later showed that it is possible to time each instruction through interrupts, invalidating the assumptions on the best temporal resolution available to the attacker. Therefore, successive defenses~\cite{branchshadowmitig} relied upon randomizing control-flow through unconditional jumps to protect enclaves.

The current understanding of the attacker's capabilities leaves the impression that as long as branches do not have observable timing differences, do not leave a different cache trace, and BPU attacks are prevented, controlled channels can be contained. As shown in the snippet of code in Listing~\ref{lst:mbedtlscost}, even widely used crypto libraries tend to use balanced branches\footnote{branches that contain the very same instructions on both execution paths} to ``prevent timing attacks.''
\lstset{basicstyle=\ttfamily\small,breaklines=true}
\begin{lstlisting}[language={C},caption={Protection against timing attacks in the latest version (v2.16.6 at the time of writing) of MbedTLS. The library balances branches by having symmetric execution paths.},label={lst:mbedtlscost},captionpos=b,float,xleftmargin=0em,belowskip=0em]
static int mpi_montmul( ... ) {
    ...
    if( mbedtls_mpi_cmp_abs( A, N ) >= 0 )
        mpi_sub_hlp( n, N->p, A->p );
    else
        /* prevent timing attacks */
        mpi_sub_hlp( n, A->p, T->p );
    return( 0 );
}
\end{lstlisting}
This might seem reasonable; after all, the branches in Listing~\ref{lst:mbedtlscost} would neither be observable with page attacks, since the same function is called on both paths, nor with Nemesis as both paths have the same instructions.
We question this last line of defense by increasing the attacker's resolution further and demonstrating that virtually any binary with control-flow secret dependencies leaks information in SGX.

\noindent\textbf{\attackname attack:}
For the first time, we show that when interrupts are frequently issued, instructions' execution time is correlated to their virtual address and that the fetch and pre-decode module of the CPU frontend plays a role in this correlation.  %
Based on this observation, we construct a new attack against Intel SGX that we call the \emph{\attackname} attack. Our attack allows an attacker to associate a measured instruction's execution time with its offset in the \emph{instruction fetch window} and thus with the instruction's virtual address. The attacker can then use these leaked execution times and addresses to infer control-flow and, therefore, branch-dependent secrets. %

We focus on extracting branch-dependent secrets, showing that an adversary can distinguish between two code sequences executed within SGX and hence, derive the secret branch condition that led to their execution. Unlike previous attacks~\cite{van2018nemesis,moghimi2020copycat}, which could only distinguish between sequences of different instructions, the \attackname attack allows the adversary to distinguish between two execution sequences even if they contain identical instructions (and even identical data). These differences are observable even when the two snippets of code reside in the same cache line and are thus not susceptible to cache side-channel attacks.
We show that by using the \attackname attack, the adversary can extract the correct secret from the enclave with probability up to $99\%$ on our test binaries.
We discuss how two different libraries, the mbedTLS library~\cite{mbed-tls} and the Intel IPP~\cite{intelipp} Cryptography library, can be exploited using our attack. Showing that, with just a few runs, the attacker can recover the condition of the executed victim branches with high confidence (> $99.9\%$), and that with a single trace it is possible to recover a full RSA key within seconds on 65\% of the runs (out of $1000$).
We validated our attack on all available Intel microarchitectures since the introduction of Intel SGX (up to Comet Lake at the time of writing).%
We show that the attack works with high probability on all tested CPUs irrespective of their microcode version. We further discuss which system configurations are better than others for the attacker. For instance, unlike in most other microarchitectural attacks, disabling hyperthreading helps the attacker.

\noindent\textbf{Defenses:}
Given the resolution achieved with our attack, a more realistic SGX adversary model should be one that considers the instruction pointer to be available to the attacker at any time. Confidentiality in SGX can only be guaranteed in this model if secret-dependent branching is avoided altogether, for instance, by if-conversion~\cite{practicalmiticoppens09} or by writing code following data-oblivious practices~\cite{sidechannelbestpractices}. These defenses are effective against any side-channel attack - including ours. However, practically deploying them is not straightforward for two reasons. First, general compiler transformations incur high-performance overheads or require developer assistance to mark secrets~\cite{practicalmiticoppens09}. Second, custom data oblivious solutions are not trivial to develop correctly and require domain-specific knowledge~\cite{sidechannelbestpractices}.

These practical hurdles for data-oblivious code have led to several spot defenses being continuously refined based on the adversary's capabilities. We give further evidence in this paper that these are bound to be broken whenever previous assumptions about the attacker are challenged.

In summary, we make the following contributions: 
\begin{itemize}
    \item We investigate how frequent interrupts affect instruction execution times. In particular, we show a dependency between the observed execution times and their alignments within the CPU fetch window.
    \item We introduce the \emph{\attackname} attack. It leverages the dependency between execution time and virtual address to attack Intel SGX enclaves. The \attackname attack leaks fine-grained control-flow in branches containing the same instructions, and that only span a single cacheline. It can do so with more than 99\% accuracy in our synthetic binaries.
    \item We exploit two commonly used cryptographic libraries using the \attackname attack: the Intel IPP Cryptography library, and the mbedTLS library. We further test which CPUs are vulnerable to our attack and found that all available CPUs with SGX at the time of writing (up to 10th gen) are vulnerable. Newer CPUs that include hardware mitigations against Spectre~\cite{spectre} seem to be more vulnerable than older CPUs. We responsibly disclosed the findings to the affected vendors.
\end{itemize}

\section{Background}\label{sec:background}

\paragraph{SGX-Step \& Nemesis}
SGX-Step~\cite{van2017sgx} is an open-source framework that allows single-stepping through the execution of SGX enclaves. SGX-Step uses APIC timers to interrupt the enclave after every instruction and inserts custom routines in between the interrupt handler and the enclave resumption.
It does not rely on any adversarial capability not given in the standard Intel SGX attacker model as interrupt handlers and APIC timers are controlled by the OS, which is assumed to be under the control of the adversary.

When an enclave receives an interrupt, it performs an Asynchronous Enclave Exit (\texttt{AEX}) and then jumps to the interrupt-vector entry defined in the interrupt descriptor table (IDT) to handle the interrupt. After the interrupt has been handled, it jumps to the address set in the asynchronous enclave pointer (AEP). The function in the AEP eventually executes the \texttt{ERESUME} instruction to resume the enclave~\cite{sgxexplained}. SGX-Step installs a custom interrupt handler in user-space to gain control as soon as possible after the interrupt. It also replaces the AEP to execute custom instructions right before \texttt{ERESUME}. SGX-Step uses these modified routines to store the current cycle count just before entering the enclave and right after an AEX. To interrupt the enclave at the right time, it configures a cycle-accurate APIC timer. This timer can be configured so that the execution is interrupted after a single instruction is executed inside the enclave. These changes allow an adversary to single-step an enclave and measure the execution time of individual instructions (including a constant offset by the \texttt{ERESUME} and \texttt{AEX}).

The Nemesis~\cite{van2018nemesis} attack exploits the fact that the interrupt timings obtained through SGX-Step are correlated with the instruction type currently pending in the CPU. 
Since current processors execute some instructions faster than others, the adversary can make an educated guess about the type of instruction that was executed in a single step. Based on a trace of these timings and knowledge of the binary executing in the enclave, the attacker can detect where the instruction pointer (IP) was in the enclave when the interrupt was received. Because Nemesis can only infer the instruction type, it cannot resolve the IP whenever a balanced branch is executed in the enclave.

\paragraph{CPU Background: The Frontend}
Although the x86 instruction set architecture (ISA) is well specified~\cite{intelSDM}, the microarchitecture is typically proprietary, and its details are confidential. 
Generally, the processor core can be split into three main parts: the frontend, the backend, and the memory subsystem. Here, we will focus on the frontend of the processor. For further information into the other components, we refer to~\cite{microarchoptfog}.

The frontend of a processor is responsible for fetching and decoding instructions into a format that the backend understands. Modern Intel processors need to fetch a large number of macro-ops to feed the extremely performant out-of-order backend. A modern Intel core fetches 16 bytes at once~\cite{microarchoptfog}, from 16 bytes aligned blocks, also called the instruction \emph{fetch window}. In x86, there is an extra step during decoding where the fetched x86 instructions (macro-ops) get translated to a different internal instruction format called micro-operations (micro-ops).

\section{Overview of the \attackname attack}\label{sec:overview}

\paragraph{Attacker model}
We consider an attacker that wants to leak secret data from a victim SGX enclave running on a system under their control. The victim enclave has a control-flow dependency related to the secret data the attacker wants to leak. The adversary operates under the standard SGX attacker model~\cite{sgxexplained}. That is, they control the entire software stack, including the operating system (OS), on the machine in which the enclave executes. Since the attacker controls the OS, we assume they can disable any CPU core to reduce noise or prevent the scheduler from running tasks on a particular core.
However, the CPU package is not physically compromised. 
We assume that the secret that the enclave holds was remotely loaded after a successful attestation. Otherwise, if the secret would be contained in the enclave code, it would be trivially available to the OS.

\paragraph{Attack overview}
We introduce our attack through an example code snippet that we show in Figure~\ref{fig:hard-c-code} (C code), and Figure~\ref{fig:asm_code} (x86 assembly). The code fits in a single cacheline and has a branch whose target depends on a secret value.
On both branches, the code contains the very same instructions and writes to the same memory addresses. Thus, we expect its execution time to be independent of which branch is taken and hence not to have any correlation with the secret input.%

\lstset{aboveskip=0pt,belowskip=-1pt}
\begin{figure}[tbp]
    \centering
    \subfloat[Secret-dependent branch\label{fig:hard-c-code}]{\resizebox{0.6\linewidth}{!}{\begin{tikzpicture}[
    code/.style={rectangle, draw=white!100,inner sep=1pt,minimum width=0.8cm,text width=4.95cm}
]
    \definecolor{cbred}{HTML} {BB5566}
    \definecolor{cbblue}{HTML} {004488}
    \definecolor{cbyellow}{HTML} {DDAA33}
    
    \node[code,align=left] (addr2) {
        \begin{lstlisting}[language=C]
if (secret == 'a') {
        \end{lstlisting}
        \begin{lstlisting}[language=C,backgroundcolor=\color{cbblue!30}]
    var1 = 1 + var1;
    var2 = 1 + var2;
        \end{lstlisting}
        \begin{lstlisting}[language=C]
} else {
        \end{lstlisting}
        \begin{lstlisting}[language=C,backgroundcolor=\color{cbyellow!30}]
    var1 = 2 + var1;
    var2 = 2 + var2;
        \end{lstlisting}
        \begin{lstlisting}[language=C]
}
return;
            \end{lstlisting}
        };
        \node[draw=none] at (3.3,0) {\color{white}{H}};
\end{tikzpicture}%
}}\\
    \vspace{-1em}
    \subfloat[Secret-dependent branch in asm\label{fig:asm_code}]{\resizebox{0.7\linewidth}{!}{\begin{tikzpicture}[
    code/.style={rectangle, draw=white!100,inner sep=1pt,text width=5.0cm,minimum height=1cm, align=left, anchor=south west}
]
    \definecolor{cbred}{HTML} {BB5566}
    \definecolor{cbblue}{HTML} {004488}
    \definecolor{cbyellow}{HTML} {DDAA33}

    \node[code] (code) {
        \begin{lstlisting}[language={[x86masm]Assembler}]
0x3:    mov (var1),  %rax
0x8:    mov (var2),  %rbx
0xc:    cmp (secret), 'a'
0xe:    jnz .else
        \end{lstlisting}
        \begin{lstlisting}[language={[x86masm]Assembler},backgroundcolor=\color{cbblue!30}]
0x10:   add $1, %rax
0x14:   mov %rax, (var1)
0x19:   add $1, %rbx
0x1d:   mov %rbx, (var2)
0x22:   ret
        \end{lstlisting}
        \begin{lstlisting}
...
        \end{lstlisting}
       \begin{lstlisting}[language={[x86masm]Assembler},backgroundcolor=\color{cbyellow!30}]
.else:
0x2b:   add $2, %rax
0x2f:   mov %rax, (var1)
0x34:   add $2, %rbx
0x38:   mov %rbx, (var2)
0x3d:   ret
        \end{lstlisting}
    };
    \coordinate (w1s) at ($(code.north west)-(0,0.2)$);
    \coordinate (w1e) at ($(w1s)-(0,1.40)$);
    \coordinate (w2s) at ($(w1e)-(0,0.15)$);
    \coordinate (w2e) at ($(w2s)-(0,1.40)$);
    \coordinate (w3s) at ($(w2e)-(0,0.15)$);
    \coordinate (w3e) at ($(w3s)-(0,1.70)$);
    \coordinate (w4s) at ($(w3e)-(0,0.15)$);
    \coordinate (w4e) at ($(w4s)-(0,0.95)$);
    \path[-,thick,draw=black!100] (w1s) -- ($(w1s)-(0.08,0)$) -- ($(w1e)-(0.08,0)$) -- (w1e);
    \path[-,thick,draw=black!100] (w2s) -- ($(w2s)-(0.08,0)$) -- ($(w2e)-(0.08,0)$) -- (w2e);
    \path[-,thick,draw=black!100] (w3s) -- ($(w3s)-(0.08,0)$) -- ($(w3e)-(0.08,0)$) -- (w3e);
    \path[-,thick,draw=black!100] (w4s) -- ($(w4s)-(0.08,0)$) -- ($(w4e)-(0.08,0)$) -- (w4e);
    \node [left =0.3cm of code,anchor=center,rotate=90] {\small{Fetch Window (16 Bytes)}};
    
    \coordinate (i1if) at ($(code.north west)-(-5.5,1.85)$);
    \coordinate (i2if) at ($(i1if)-(0,0.39)$);
    \coordinate (i3if) at ($(i2if)-(0,0.39)$);
    \coordinate (i4if) at ($(i3if)-(0,0.39)$);
    \path[->,thick,draw=black!100,>=stealth] (i1if) -- ($(i1if)-(0.4,0)$);
    \path[->,thick,draw=black!100,>=stealth] (i2if) -- ($(i2if)-(0.4,0)$);
    \path[->,thick,draw=black!100,>=stealth] (i3if) -- ($(i3if)-(0.4,0)$);
    \path[->,thick,draw=black!100,>=stealth] (i4if) -- ($(i4if)-(0.4,0)$);
    \node[draw=none] at ($(i1if)+(0.5,0)$) {Int \#1};
    \node[draw=none] at ($(i2if)+(0.5,0)$) {Int \#2};
    \node[draw=none] at ($(i3if)+(0.5,0)$) {Int \#3};
    \node[draw=none] at ($(i4if)+(0.5,0)$) {Int \#4};

    \coordinate (i1else) at ($(code.north west)-(-5.5,4.50)$);
    \coordinate (i2else) at ($(i1else)-(0,0.39)$);
    \coordinate (i3else) at ($(i2else)-(0,0.39)$);
    \coordinate (i4else) at ($(i3else)-(0,0.39)$);
    \path[->,thick,draw=black!100,>=stealth] (i1else) -- ($(i1else)-(0.4,0)$);
    \path[->,thick,draw=black!100,>=stealth] (i2else) -- ($(i2else)-(0.4,0)$);
    \path[->,thick,draw=black!100,>=stealth] (i3else) -- ($(i3else)-(0.4,0)$);
    \path[->,thick,draw=black!100,>=stealth] (i4else) -- ($(i4else)-(0.4,0)$);
    \node[draw=none] at ($(i1else)+(0.5,0)$) {Int \#1};
    \node[draw=none] at ($(i2else)+(0.5,0)$) {Int \#2};
    \node[draw=none] at ($(i3else)+(0.5,0)$) {Int \#3};
    \node[draw=none] at ($(i4else)+(0.5,0)$) {Int \#4};

\end{tikzpicture}%
}}
    \caption{A secret-dependent branch in C and x86 assembly. Both branches in the assembly code fit within the same cacheline (64B). The virtual address of the instructions is reported on the left. Note that while the branches are instruction-wise identical, their instructions get grouped differently by the fetch window (which always start at multiples of 16B).}
    \label{fig:code}
    \vspace{-1em}
\end{figure}

However, when the above sequence is run within an SGX enclave, our attack shows that a local attacker can learn which branch was taken, and therefore, derive the secret value of the branch condition. 
Our attack leverages two main observations. First, even if the branches have the same instructions, they are often aligned differently within the fetch windows  (Listing~\ref{fig:asm_code}) -- in our experiments, this alone did not produce observable differences in the execution times (cf.\ Section~\ref{sec:obs}). Second, if the execution of both branches is frequently interrupted, the difference in their alignments w.r.t.\ the fetch windows will cause the CPU to fetch instructions at different times (Table~\ref{tab:aq_interrupts}), resulting in a measurable difference in the execution times of the instructions and therefore of the branches (cf.\ Section~\ref{sec:obs}).

To give an insight into why interrupts lead to a successful attack, we show which instructions are fetched by the CPU when the execution is interrupted after each instruction. There are two main factors to consider: which instructions among those already in the pipeline are retired when an interrupt is received, and how execution is resumed after an interrupt.
Intel guarantees that only the oldest pending instruction in the reorder buffer is retired~\footnote{Or discarded, if it raises an exception} before the interrupt is handled~\cite{van2017sgx}. In out-of-order processors, other instructions might have already been executed, but none of these will be retired.
To resume execution after the interrupt is handled, the CPU needs to fetch the instruction sequence starting at the current program counter.
However, while the program counter can, in general, have any value, fetch windows are statically aligned at 16 bytes code blocks~\cite{microarchoptfog}. Assume that the program counter falls 5 bytes after the start of the fetch window. Those initial 5 bytes will be fetched only to be then discarded by the frontend. Thus out of 16 bytes fetched, only 11 are usable. Now assume that the same instruction sequence begins 10 bytes after the start of the fetch window. Instead of 11 bytes as before, there are only 6 bytes that can be decoded, meaning we now need two fetch windows (and hence two cycles) to decode the same number of instructions as we did before in just one fetch window.
Alignment w.r.t.\ fetch windows can, therefore, change the order in which instructions are forwarded to other stages of the CPU and ultimately populate the pipeline. To help clarify this point, for both branches of our example code, we show in Table~\ref{tab:aq_interrupts} which instructions are fetched after every interrupt.

In principle, given the same system conditions, a particular instruction should exhibit the same time distribution at different virtual addresses.
However, we experimentally observe that depending on the alignment within a fetch window and the number and type of instructions present around them, some instructions consistently take longer to execute than others. In Section~\ref{sec:obs}, we provide more details on which alignments of instructions produce measurable execution time differences. 
This observation hence allows us to associate the measured instruction execution time with the alignment in the fetch window, and therefore with the instruction virtual address (i.e., with the instruction pointer). 
These leaked execution times and addresses can then be used to infer executed branches (e.g., when they depend on the secret value). %
In this work, we focus on the use of our attack in the context of secret-dependent branching. In particular, for the scenario given above in Table~\ref{tab:aq_interrupts} when enough \texttt{mov} are fetched after a \texttt{mov} in the branch, the interrupt latency is measurably different. In our example, we measured interrupt \texttt{\#2} in the table to be faster if the code is executing in the ``else'' branch, as compared to the ``if'' branch, despite the fact that we are interrupting the \emph{same instruction under the exact same system conditions}.

\begin{table}
\centering
\resizebox{0.8\linewidth}{!}{%
\begin{tabular}{@{}llllllll@{}}

\toprule
                         & \multicolumn{3}{c}{If}                                   & \multicolumn{4}{c}{Else} \\ \cmidrule(lr){2-4} \cmidrule(ll){5-8}
Int \#1                  & add &  mov  & add  & add    &        &        &   \\
\textbf{Int \#2}         & \textbf{mov}     &  \textbf{add}    &                  & \textbf{mov}    &  \textbf{add}   & \textbf{mov}    & \textbf{ret}     \\
Int \#3                  & add                &                   &                  & add    &  mov   & ret    &   \\
Int \#4                  & mov              & ret              &                  & mov    & ret    &        &   \\ \bottomrule
\end{tabular}}
\caption{Here we show how instructions are batched into fetch windows when the enclave resumes execution, according to which branch is executing. If an instruction crosses a fetch window boundary, we assume it is decoded together with the instructions in the following window. The interrupts refer to the instructions in Figure~\ref{fig:asm_code}.}\label{tab:aq_interrupts}%
\vspace{-1.0em}
\end{table}

Let's again consider Listing~\ref{fig:asm_code}. By running SGX-Step, we can time all instructions by stepping through them one by one.
As a consequence of the observations made above, we will observe two scenarios for the 6th instruction measured, which is the instruction at address 0x14 or 0x2f, depending on the secret value. If the interrupt is ``slower'' (compared to the others measured), we must be executing the \texttt{mov} at address 0x14. Inversely, if the interrupt is ``faster'', we must be executing the \texttt{mov} at address 0x2f. Since the control flow of the program depends on the secret, this allows us to recover its value, and hence break the SGX confidentiality guarantees.

The snippet presented in Figure~\ref{fig:code} produces distinguishable timings for the first \texttt{mov} instruction inside the branch. We were able to use the timing difference to predict the secret with $\ge 65\%$ accuracy. By adding three more \texttt{mov}s after the branches (which are executed by both paths), we were able to obtain success rates $>90\%$.
The attack presented above illustrates how fully balanced branches actually produce secret-dependent timings when interrupted frequently. Given that this side-channel is due to the design and behavior of the CPU frontend, we name our attack the \emph{\attackname} attack.
In the following sections, we will analyze our attack in more detail.

\section{\attackname Attack Profiling}\label{sec:obs}

In this section, we provide more detail and clarification to that help in understanding under which circumstances the \attackname attack works. More specifically, we ask and answer the following questions: (i) are the interrupts required for the attack to be successful? (ii) what are the effects of the fetch window alignment / instruction address on the attack? and (iii) which instructions produce observable timing differences? 

To answer these questions we perform experiments over the code snippet shown in Figure~\ref{fig:asm_long_code}. Similar to code in Figure~\ref{fig:code}, this code snippet contains two perfectly symmetric branches depending on a secret. It still consists of two perfectly balanced branches but differs in that now each branch contains 25 sequences of \texttt{add-mov} instructions.
We chose this longer code sequence since it produces timing differences that are more clearly above the noise floor than the code in Figure~\ref{fig:code} and therefore better illustrates timing and alignment effects under different experiment configurations. Namely, code sequences that include several \texttt{mov} instructions like the one in Figure~\ref{fig:asm_long_code} are particularly susceptible to the \attackname attack and allow us to extract the secret branch condition with an accuracy of at least $99\%$, whereas with shorter sequences that contain few \texttt{mov}s (like the one in Figure~\ref{fig:code}), this accuracy drops to $\ge 65\%$. We discuss this effect in more detail later in this section.

\begin{figure}[tbp]
    \centering
    \resizebox{0.6\linewidth}{!}{%
        \begin{tikzpicture}[
    code/.style={rectangle, draw=white!100,inner sep=1pt,text width=5.5cm,minimum height=1cm, align=left, anchor=south west}
]
    \definecolor{cbred}{HTML} {BB5566}
    \definecolor{cbblue}{HTML} {004488}
    \definecolor{cbyellow}{HTML} {DDAA33}
    
    \node[code] (code) {
        \begin{lstlisting}[language={[x86masm]Assembler}]
        .align (x - 0x4)
x - 0x4:    cmp (secret), 1
x - 0x2:    jnz .else
        \end{lstlisting}
        \begin{lstlisting}[language={[x86masm]Assembler},backgroundcolor=\color{cbblue!30}]
        .if:
        .rept 25
x + 0x0:    add %rax, %rbx
x + 0x3:    mov %rcx, (var1)
        .endr
x + 0x190:  ret
        \end{lstlisting}
        \begin{lstlisting}[language={[x86masm]Assembler}]
...
        .align y
        \end{lstlisting}
        \begin{lstlisting}[language={[x86masm]Assembler},backgroundcolor=\color{cbyellow!30}]
        .else:
        .rept 25
y + 0x0:    add %rax, %rbx
y + 0x3:    mov %rcx, (var1)
        .endr
y + 0x190:  ret
        \end{lstlisting}
    };
\end{tikzpicture}
    }
    \vspace{-4pt}
    \caption{ASM Code with high attack success probability, which we use to profile the attack. The \texttt{.rept 25 ... .endr} assembler directive repeats the instructions within the block 25 times, leading to an address of \texttt{x+0x190} for the \texttt{ret} instruction.}\label{fig:asm_long_code}
\end{figure}

\subsection{The Role of Interrupts}\label{sub:interrupts}

To analyze the effect of frequent interrupts on the behavior of the processor we measure the execution time of our test code snippet (Figure~\ref{fig:asm_long_code}) with and without interrupts. %

\begin{figure}[tbp]
    \centering
    \includegraphics[trim={16 16 16 16},clip,width=\linewidth]{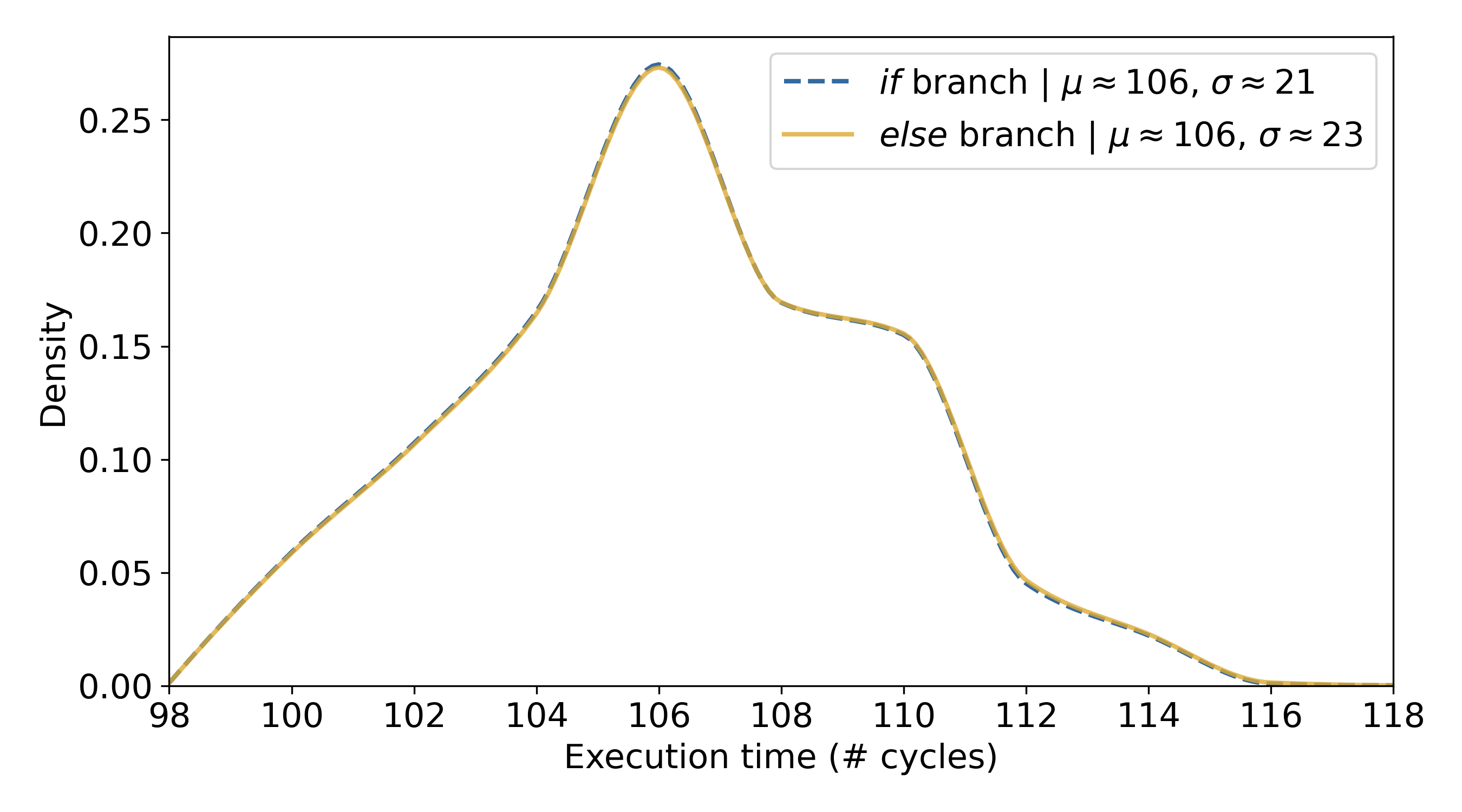}
    \vspace{-10pt}
    \caption{Distribution of the overall execution time of the branches in Figure~\ref{fig:asm_long_code} when run outside of SGX without interrupts (computed from $2 * 10^9$ samples).}
    \label{fig:outside_distr}
\end{figure}

\paragraph{Outside SGX without interrupts}
We first measured the overall execution time of the code snippet outside SGX without interrupts. We executed the code with two billion independent random inputs, and we observed no significant correlation between execution times and the branch that was executed (Pearson's coefficient $= -2.51 \cdot 10^{-5}$). An approximate distribution of this measurement is shown in Figure~\ref{fig:outside_distr}.

\paragraph{In SGX without interrupts}
In order to exclude any effect due to SGX, we further measure the overall execution time of the code within an SGX enclave, again without interrupts. Note that SGX does not provide any way to get a precise timer (cf.\ Section~\ref{sec:background}), so we have to measure the execution time from the untrusted app. 

We perform this measurement using three different methods. 
All methods use the same code snippet in a loop, but they differ in how the measurement is collected and where the loop is executed. We do this to filter out any effects of the enclave entry and exit operations.
First, we measure a whole enclave call from the untrusted app. Multiple measurements are collected by having a loop in the untrusted app.
Second, we run the loop entirely inside the enclave and collect the iteration execution time with two \texttt{ocall}s to the untrusted app. The two \texttt{ocall}s are done at the beginning and the end of each loop iteration.
Third, we use a similar setup as the second method, but instead of performing \texttt{ocall}s, the enclave samples the value of a counter stored in shared memory. A thread of the untrusted app increments the counter in a loop, thus simulating the time stamp counter, albeit at a lower precision. All three methods use an independent uniform random value as the ``secret'' given to the code at each iteration.

For all three methods, similar to the experiment outside of SGX, we observed no significant correlation (Pearson's coefficient $\approx10^{-2}$ with $10^6$ runs) between the execution time and the secret provided to the enclave. %

\paragraph{In SGX with interrupts}
We now investigate which effects frequent interrupts have on the execution time of the code. We execute the same code snippet as before but we interrupt it after each instruction. Upon each interrupt the CPU performs an asynchronous enclave exit (AEX), handles the interrupt, and then performs an \texttt{ERESUME} to resume the enclave execution. 
Such an experiment would normally require very fast and extremely precise interrupts, which is usually hard to achieve. However, in the case of a victim code running within SGX, we can use SGX-Step~\cite{van2017sgx} to single step through each instruction and collect its execution time. 
Given these interrupts, we can not only measure the overall execution time, but also the execution time of each instruction. This means that in each run of our code, we obtain 51 measurements.\footnote{There are 52 instructions in Figure~\ref{fig:asm_long_code}, however the first \texttt{cmp} and \texttt{jnz} get macro-fused into one instruction which cannot be split again by interrupts.}

We then analyzed whether any of the 51 measured instruction execution times correlate with the executed branch. We observed a \emph{strong correlation} between the timings of most of the instructions and the branch they belong to. The first 10 \texttt{mov} instructions in the branch turned out to be a stronger indicator of which branch was taken, but all the other instructions belonging to the branch showed some correlation, albeit a weaker one.\footnote{The timings of the initial \texttt{cmp} and \texttt{jnz} were independent of the executed branch - only instructions within the branches were correlated with the secret.}

As in Section~\ref{sec:overview}, we observed the execution time of the first \texttt{mov} in each branch to be faster or slower, depending on the branch it belongs to, with a difference between the slower and faster \texttt{mov} of around 100 cycles. %
This observation allowed us to set a timing threshold with which we could, with up to 99.9\% accuracy, determine which branch was taken, and therefore determine the secret branch condition.%

We stress again that the two branches are instruction-wise identical: the instructions they contain \emph{and their inputs} are the same. This is especially important because it highlights the fact that the timing difference is due to the way the instructions are executed, and not some external system state. For instance, the difference \emph{cannot} be due to the state of the cache, the state of the branch predictor, or in general to some speculation decisions made by the CPU. If the cause of the differences were to be due to any of these factors, we would expect two key differences. First, as we choose secrets at random, these effects would manifest with equal probability in any of the two branches. Second, we would expect the experiments in which we do not interrupt the code to also show some bias. However, we see a clear bias in one of the two branches, and the interrupt-free runs showed no correlation with the secret.

\begin{ObservationBox}{1}\label{obs:int}
When code execution is frequently interrupted, the execution times of selected instructions depend on their location in the victim binary and therefore on their virtual memory address.
\end{ObservationBox}

\subsection{Relationship to Virtual Addresses}

While the instructions in both branches are identical, there is one key difference between them: their virtual address. Therefore, we analyze what virtual addresses make the two branches distinguishable when frequently interrupted, and to what degree. %
In particular, as discussed in Section~\ref{sec:overview}, we also study how the relationship between the alignment of the branches with respect to the fetch window affects the success of the attack.
As can be seen in Figure~\ref{fig:asm_long_code}, we use the \texttt{align} compiler directive to explicitly align each branch to a given address. 
With \texttt{.align X} we indicate that the code following the directive starts at the next virtual address whose lower bits are equal to \texttt{X}.\footnote{This is equivalent to combining the two gcc asm directives \texttt{.align $(X // 2^n)$} and \texttt{.space $(X \% 2^n)$} (for the biggest $n$ such that $2^n < X$)} For example, if $X = 3$ and $Y = 2$, then the \texttt{if} branch will start at address \texttt{0x13} and end at address \texttt{0x1a3}, while the \texttt{else} branch will start at address \texttt{0x1b2}.%

\begin{figure}
    \centering
    \includegraphics[width=0.9\linewidth]{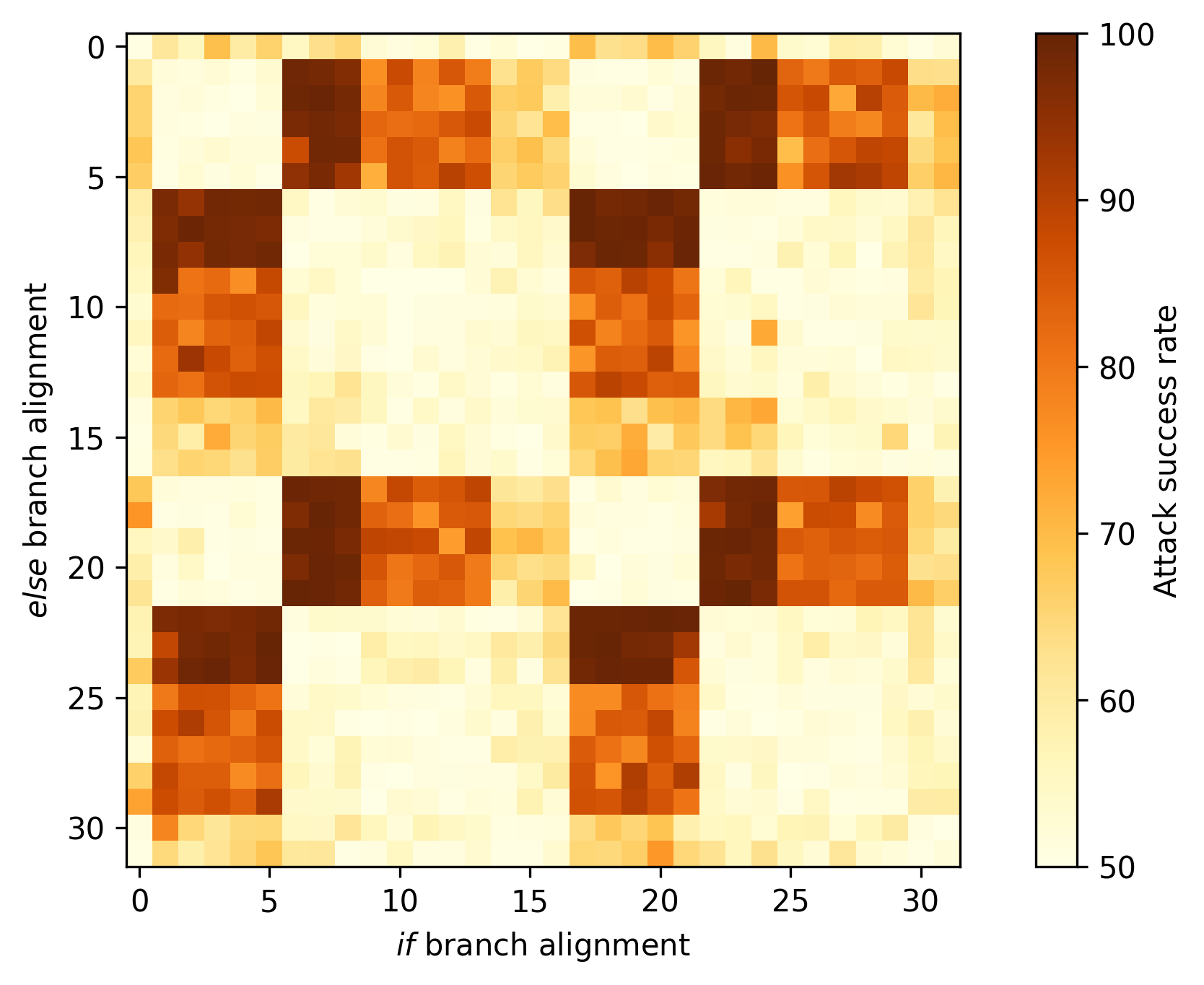}
    \vspace{-8pt}
    \caption{Attack success rate depending on the alignment of the branches. The attack success rate is the percentage of correctly guessed branches by the attacker out of 1000 executed branches. The 10th instruction (5th \texttt{mov}) from Figure~\ref{fig:asm_long_code} is used to distinguish between both branches. The color gradient goes from darker to brighter, where darker boxes indicate higher attack success rates (up to 100\%) and brighter ones lower success rates (down to 50\%).}
    \label{fig:heatmap}
    \vspace{-0.5em}
\end{figure}

To evaluate different alignments, we run an experiment to test if different values of \texttt{X} and \texttt{Y} in Figure~\ref{fig:asm_long_code} have any effect on the observed timing differences. We repeat the interrupt experiment described at the end of Section~\ref{sub:interrupts}. That is, we send an interrupt to each instruction and then use the interrupt timing of one of the instructions in the branch as a discriminator to determine which branch was taken, and thus what the secret was. We then calculate the attack success as the percentage of correctly identified secret bits. Therefore, the attack success rate will tell us how good a certain combination of the alignments \texttt{X} and \texttt{Y} are for the attack. 
The higher the percentage the better an alignment combination is for the attack, while a result close to 50\% indicates that predicting which branch was taken is as good as a random guess.
We collect these percentages for each combination of $\lbrace X, Y\rbrace \in [0, 31]^2$ by running the code in Figure~\ref{fig:asm_long_code} $1000$ times with uniformly random secrets. We use the timings of the 10th instruction (5th \texttt{mov}) to discriminate between the branches.
Figure~\ref{fig:heatmap} presents the result of our experiment. These results show a clear dependency between virtual addresses and the instruction execution times.

\paragraph{Modulo 16}
There are four main quadrants of length $16$ that are essentially identical. 
This hints at the fact that the behavior with respect to the alignment of the two branches repeats every 16 bytes. We verified this assumption by repeating the experiment for every value of \texttt{X} and \texttt{Y} for which the two branches are still contained in the same 4 kB virtual page.\footnote{We did not cross the virtual page boundary because this would most likely require fetching pages that are not cached, thus introducing noise that masks the effects that we are interested in measuring.} We observed the same pattern for all the quadrants of length $16$ in this test. As a consequence of this observation, when we use the term alignment, we refer to alignment modulo 16.

\begin{ObservationBox}{2.1}\label{obs:mod16}
The attack success rate depends on the alignment \emph{modulo 16} of the two branches.
\end{ObservationBox}

\paragraph{Diagonals}
The attack success rate on the diagonals in each quadrant is around 50\%. In the diagonals, both branches are aligned to the same value \texttt{X = Y mod 16}. %

\begin{ObservationBox}{2.2}\label{obs:same_al}
Branches and instructions with the same alignment will show the same execution times.
\end{ObservationBox}

\paragraph{Symmetry}
The attack success rates are symmetric with respect to their diagonal, meaning that the success of the attack when the ``if'' branch is aligned at address \texttt{X} and the ``else'' branch at address \texttt{Y} is the same when the alignment of the branches is switched.

\begin{ObservationBox}{2.3}\label{obs:comm}
Alignments $X, Y$ and $Y, X$ produce the same attack success rate. 
\end{ObservationBox}

\paragraph{Shape}
Finally, we focus our attention on the alignments in the heatmap in which the success rate is above ~70\%.
These success rates are grouped into rectangles. Within each of these rectangles, there are three regions of decreasing intensity.
The most interesting alignments are the ones that give the higher attack success rates, as they allow to optimize the accuracy of the attack. %
The best results are concentrated on rectangles of size $3 \times 5$. This corresponds with the length in bytes of the two instructions within the branch in Figure~\ref{fig:asm_long_code}. The \texttt{add} instruction has a length of $3B$, while the \texttt{mov} we use in Figure~\ref{fig:asm_long_code} has a length of $5B$. Unfortunately, this rule does not trivially generalize with more complex instruction size combinations.

Note that there are only a few structures in the CPU that are sensitive to the alignment of the instruction, and in particular, to their alignment modulo 16. On Skylake and Coffee Lake architectures, one of them is the \emph{instruction pre-decode and fetch} module in the frontend of the CPU, which uses a fetch window of 16 bytes to fetch instruction from the L1 instruction cache. 
We cannot be entirely sure about the internal behavior of the CPU and what leads to the timing differences in the two branches. However, as discussed in Section~\ref{sec:overview}, the different alignment changes the way instructions are batched by the frontend and, ultimately, the timing at which they are delivered to the subsequent stages of the CPUs. The experiments presented in this section strongly suggest that these fetching differences have repercussions for the instruction's execution time. We will discuss potential causes that could lead the observed variable timings in Section~\ref{sec:causes}.

\begin{figure}[!t]%
    \centering
    \includegraphics[trim={16 16 16 14},clip,width=\linewidth]{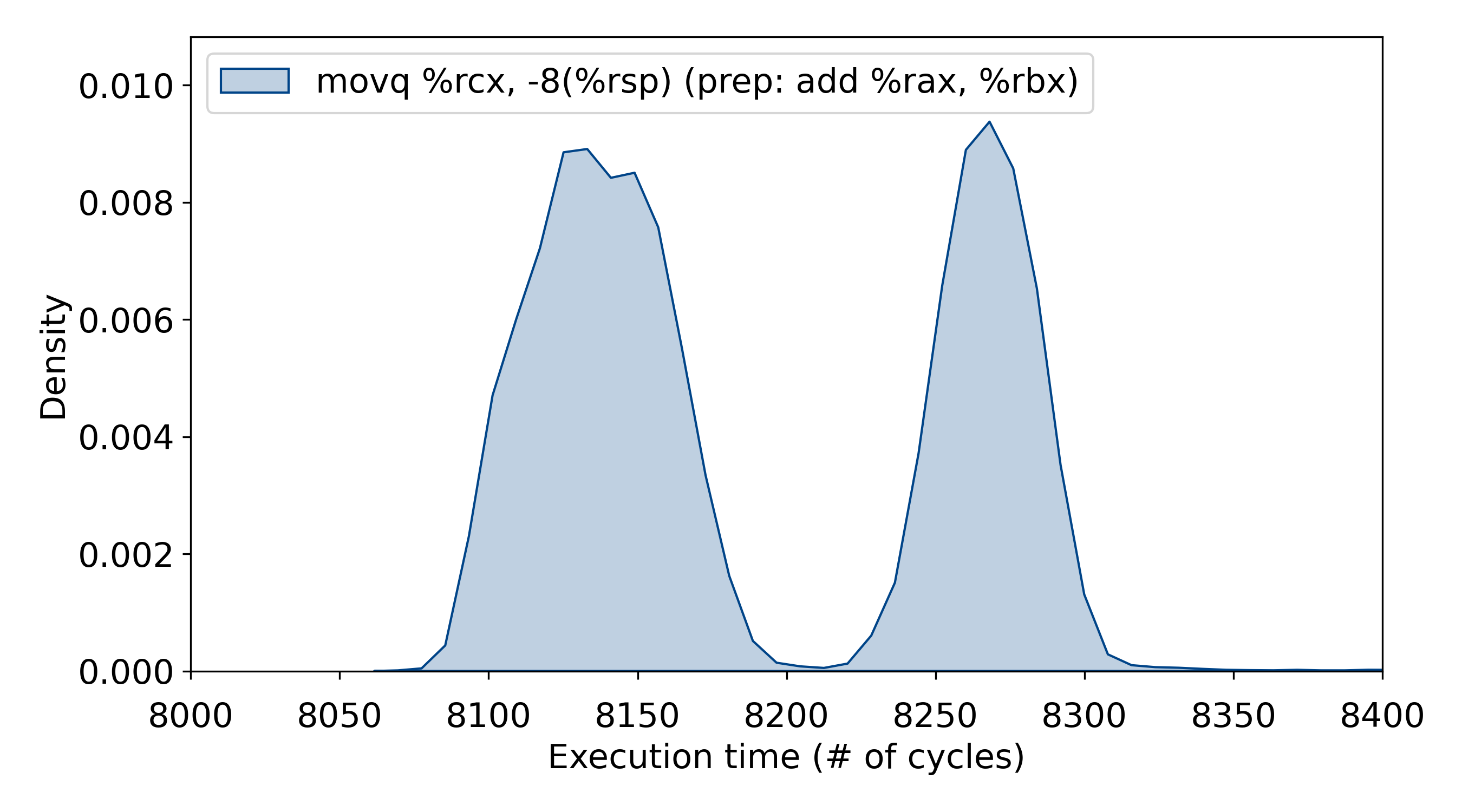}
    \vspace{-8pt}
    \caption{Timing distribution of a \texttt{mov} to the stack when executing it in a trace containing $100,000$ repeated \texttt{add-mov} instructions (unrolled).}
    \label{fig:long_seq_glob}
\end{figure}

\subsection{The Effects of Instruction Alignment}\label{sec:align}
To study the effects of the instruction alignment we analyze the timing distributions of a linear code sequence of 100,000 repeating \texttt{add-mov}. Note that this is essentially an unrolled loop, which compared to a loop removes the noise that the loop-control instructions would introduce.
We don't envision any real code to have such a sequence of instructions, but by exploring the patterns that emerge from these instructions we can gather several insights about how the differences in branch alignments manifest.

\begin{figure}[!t]%
    \centering
    \includegraphics[trim={16 16 16 14},clip,width=\linewidth]{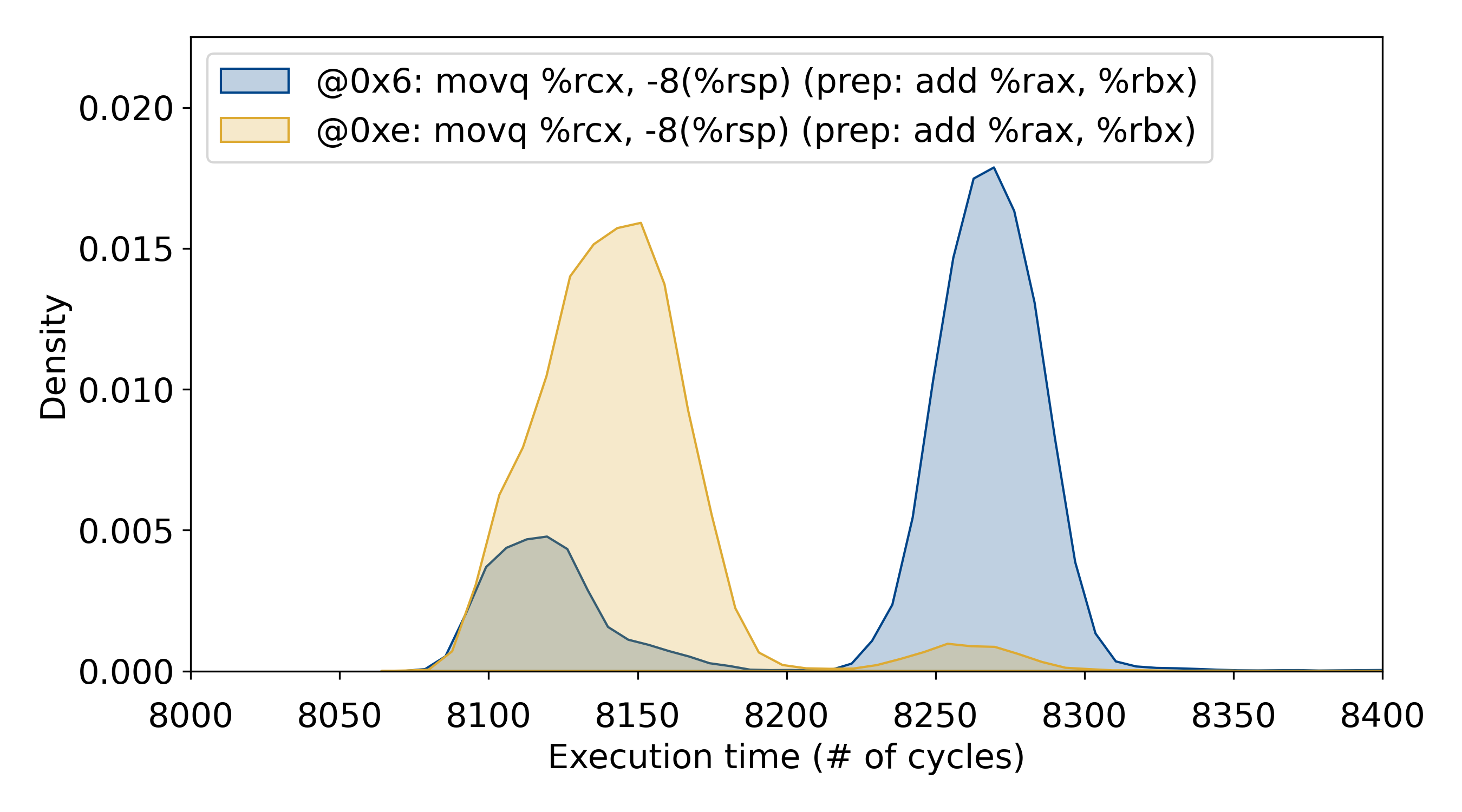}
    \caption{Timing distribution of the \texttt{mov}s from Figure~\ref{fig:long_seq_glob} grouped by their virtual address alignment.}
    \label{fig:long_seq_split}
\end{figure}

The timings are collected using a slightly modified version of SGX-Step, whose changes are described in Appendix~\ref{sec:setup}.
The timing of each instruction includes the time to perform \texttt{ERESUME}, the time to execute the instruction, and the time required to perform \texttt{AEX}. \texttt{ERESUME}, and \texttt{AEX} prepare the CPU for the enclave execution and clean the state when returning to the untrusted app. These operations take thousands of CPU cycles to complete, and this is why, despite the fact that we are measuring a single instruction, the latencies reported in the graphs are in the order of thousands of cycles.
We use two figures to illustrate different aspects of the timing latency of the same run: (i) Figure~\ref{fig:long_seq_glob} depicts the overall latency distribution of all the \texttt{mov}s, and (ii) Figure~\ref{fig:long_seq_split} the distribution separated by particular virtual addresses. %

\paragraph{Distribution of instruction execution times}
In Figure~\ref{fig:long_seq_glob} we present the distribution of the instruction execution times, estimated from all the 100,000 executed \texttt{mov}. %
The most evident feature of this distribution is that it consists of a \emph{bimodal} Gaussian distribution. %
The \texttt{mov}s are therefore exhibiting two different distribution modes, whose peaks are, on average, around 100 cycles apart. We refer to the mode with the lower average and the one with the higher average as the \emph{fast mode} and \emph{slow mode}, respectively. %

\begin{ObservationBox}{3.1}\label{obs:write_fast_slow}
The timing distribution of the \texttt{mov}s follows a bimodal distribution. The peaks of the two distribution modes are around 100 cycles apart.
\end{ObservationBox}

In general, we observed similar results with other instructions that access memory, such as \texttt{add} to memory. We remark here that these differences are not due to the state of the L1 data-cache. We ensure this by running the victim enclave on a dedicated physical core in the system and by always performing the same operations while handling interrupts. We further verified with the \texttt{OFFCORE\_REQUESTS\_ALL\_REQUESTS} performance counter that no extra off-core memory transactions were being performed.

\begin{ObservationBox}{3.2}\label{obs:memory_writes}
Observation~\ref{obs:write_fast_slow} applies not only to \texttt{mov}s but to all \emph{memory writes}. 
\end{ObservationBox}

\paragraph{Instruction execution times by alignment}\label{sec:mov_alignment}

Regarding alignment, there is an important characteristic of the chosen instruction sequence that has not been considered in our analysis thus far. Each couple of \texttt{add-mov} in the sequence has a length of 8B, which is a multiple of 16. This implies that the \texttt{mov}s \emph{can only be aligned modulo 16 in two different ways}. %
In general, by testing the sequence with different initial offsets, we observed \texttt{mov}s at addresses between \texttt{1 and 8} to be predominately slow and \texttt{mov}s at addresses \texttt{9 to 16} to be predominately fast. 
We highlight that the two alignments are only \emph{predominately} fast (or slow) and usually they exhibit timings from both distribution modes. We can think of each instruction at a given alignment to have a certain intrinsic probability $p$ to exhibit the fast mode and probability $1 - p$ to exhibit the slow mode every time it executes. Different alignments have a different value of $p$.  %
Figure~\ref{fig:long_seq_split} shows this phenomenon for two particular alignments (\texttt{0x6} and \texttt{0xe}). As can be seen, alignment \texttt{0x6} is predominately slow, but some of its timings exhibit the fast mode as well. The plots for other alignments are similar, with the only difference being the size of the smaller peaks. We do not show them here due to space constraints.

\begin{ObservationBox}{3.3}\label{obs:behavealign}
The alignment of the memory writes determines how their latency will distribute between the fast and slow distribution modes. %
\end{ObservationBox}

The value of $p$ relates to the attack success rate. Say that one branch is aligned such that the measured \texttt{mov} has $p \ge 0.9$, and the other is aligned to have a $p \le 0.1$ then the branches are easily distinguishable, and a high success rate will be observed. If one of them has $0.3 \ge p \le 0.7$, and the other a very small or very high $p$, as is the case for the distributions in Figure~\ref{fig:long_seq_split}, then one bit can be distinguished with high accuracy, but the other will contain some errors. 
If the branches have a $p \approx 0.3$ and say $p \approx 0.7$, then both branches will be on average guessed better than random, but will also contain errors. And finally if both branches have a  similar $p$ the success rate of the attacker will be negligible.

\subsection{Requirements and Limitations}

In our experiments, we only observed timing differences in branches which contain \emph{memory writes}. Thus, at least a memory write must be present for the side-channel to emerge. All the other conditions being equal, other memory write instructions we tested (variations of \texttt{mov} to different addresses and arithmetic instructions that write back to memory), excluding the \texttt{push} instruction, exhibited the very same behaviors as described so far. Notably, instructions that are surrounded by other memory writes also show a timing difference, albeit usually smaller. Furthermore, the timing distribution of a memory write is not only determined by its alignment in isolation, but it is also influenced by the number and alignment of surrounding memory instructions. For instance, the more memory writes in the branch (or even right after it), the more distinguishable the distributions will be, increasing the probability of success of the attack.
Another element we were able to characterize, relates to the vicinity of the memory instructions with each other. Particularly, when writes are executed in a loop, the attack success probability is higher if the loop executes only a few instructions (around 10) in between writes, and the fewer, the better, for the attack.

It is worth noting that simultaneous multi-threading (SMT) was a big source of noise in our experiments. When the core co-located with the victim is executing a CPU-heavy workload, we were unable to observe any significant timing difference. In general, the \attackname attack is more reliable if SMT is disabled or the virtual core co-located with the victim is idle. We speculate that this is most likely due to how the frontend handles and fetches instructions coming from different virtual cores, but possibly also to the resulting lower interference in the memory subsystem.

\section{\attackname Attack Exploitation}\label{sec:real_attack}

The \attackname attack exploits control-flow secret dependencies. Therefore, the first step of the attack is to identify target code paths in the victim binary which execute secret-dependent branches. Several techniques have been proposed to automate finding such code paths~\cite{weiser2018data,microwalk}. %
Among these code paths, as discussed before, the attacker should choose those that contain at least one memory write. Until now, we mostly focused on balanced branches, but unbalanced branches are also distinguishable with our attack. As unbalanced branches can be exploited with other attacks as well, we focus on more challenging balanced branches in our example exploits below. %
Balanced branches are not rare in compiled code. In fact, we found two code patterns that commonly lead to this type of branches: slightly different return statements, and inlined function calls with different parameters.

In the following, we give examples of vulnerable branches satisfying the conditions above in two libraries: the Intel IPP Cryptography library~\cite{intelipp}, and the mbedTLS library~\cite{mbed-tls}.
We note that since a secret-dependent code path must be present, branch-prediction attacks can also exploit the binaries vulnerable to the \attackname attack. For instance, the examples we present below, when compiled with \texttt{gcc} are also vulnerable to branch-shadowing attacks~\cite{lee2017inferring}. However, when compiling the mbedTLS library with the compiler from~\cite{branchshadowmitig}, all the branches are translated to indirect unconditional jumps, which are hitherto not vulnerable to any known BPU attack. On the other hand, we verified that even when using~\cite{branchshadowmitig} the branch targets are unchanged and have in general different alignments, thus remaining vulnerable to the \attackname attack.
The attacks described in this section were performed on an Intel i9-9900KS CPU with the latest microcode available at the time of writing (\texttt{0xca}).

\subsection{Intel IPP Cryptography Library}\label{sec:ipp_intel_vuln}
\begin{figure}[t]
    \centering
    \includegraphics[trim={0.2cm 0.4cm 0.2cm 0.2cm},clip,width=0.9\linewidth]{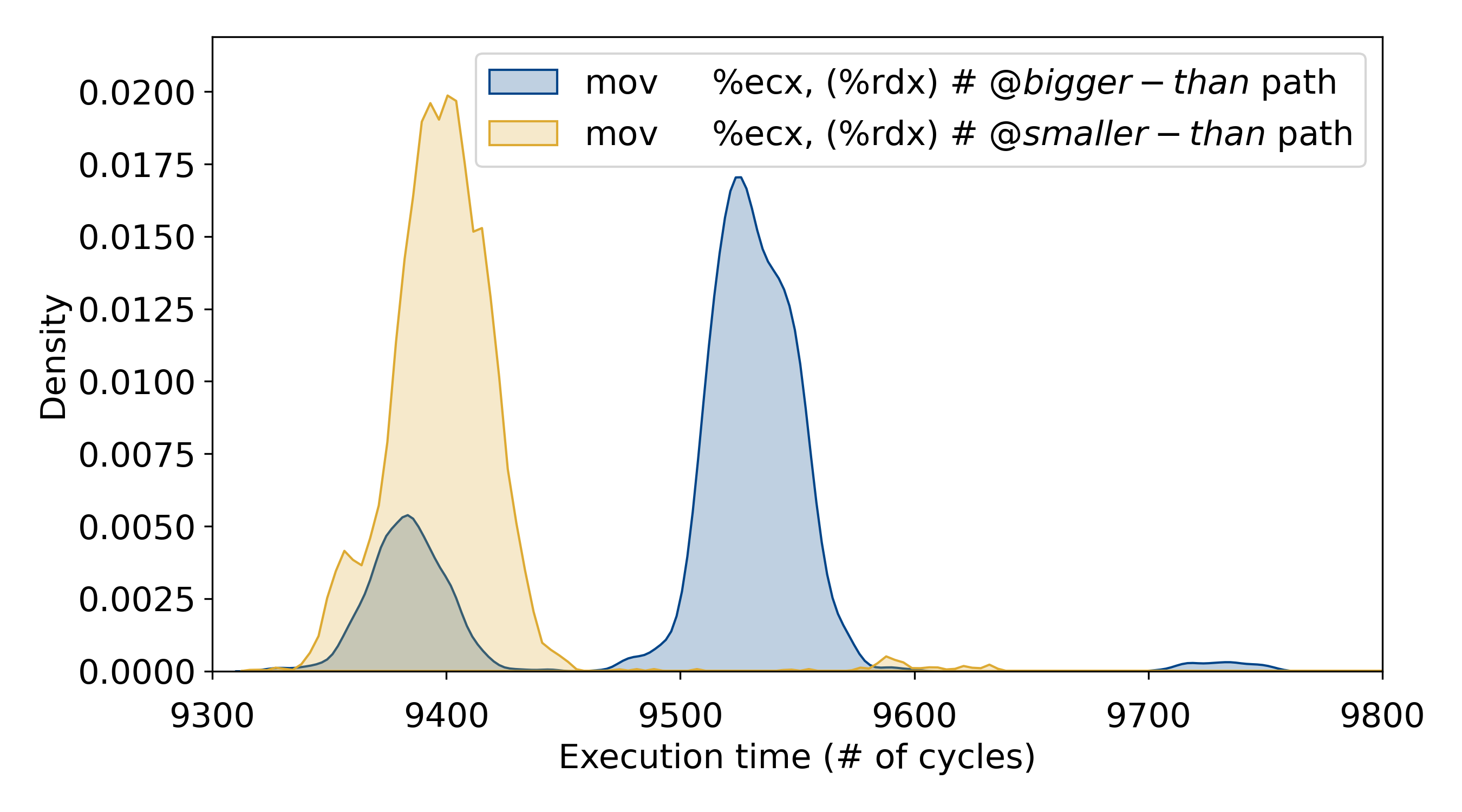}
    \caption{Timing distributions of two different \texttt{mov}s in the IPP Cryptography library's \texttt{l9\_ippsCmp\_BN} function (each estimated from $3000$ samples). The function executes a secret dependent comparison, which can result in two balanced paths being taken: the \emph{bigger-than} or \emph{smaller-than} path. Each path contains a differently-aligned \texttt{mov} in it, whose distribution is shown in the figure.}
    \label{fig:ipp_attack}
    \vspace{-1em}
\end{figure}

The Intel IPP Cryptography library is a cryptographic library optimized for Intel CPUs and advertised as constant-time~\cite{intelipp}.
However, through manual inspection we identified several secret dependent branches in its most recent version (2.9 at the time of writing). 
Among these, the \texttt{l9\_ippsCmp\_BN} function %
compares two big numbers represented as arrays of integers by iterating through each element of the array.
The function then terminates when a different array entry is found. It can take three different exit paths, depending on whether the first input is smaller, bigger, or equal to the second.
The \emph{smaller-than} and \emph{bigger-than} paths are instruction-wise identical, while the \emph{equal} path contains the same instructions as the others but in a different order. 
Given that the different order of instructions of the equal vs.\ unequal paths can be inferred with other attacks, we focus on distinguishing the \emph{smaller-than} vs.\ \emph{bigger-than} paths with the \attackname attack. With branch-prediction mitigations in place, other known attacks do not allow to leak this information, as all the paths fit in a single cache-line.
The exit paths contain a \texttt{mov} to memory, which we target in our attack. We did not observe any timing difference on this instruction alone, despite the fact that the paths start at different alignments, this is expected as the memory write is executed only once. However, by inlining the function in an enclave that performs a loop of at least 9 memory writes after the IPP function call, we obtained the distributions shown in Figure~\ref{fig:ipp_attack}.
The figure shows two distributions that differ in their modality. The timing distribution of the \texttt{mov} in the \emph{smaller-than} path has a single peak around $9400$ cycles. On the other hand, the \texttt{mov} in the \emph{bigger-than} path exhibits two modes, a small one around $9300$ cycles, and a predominant one at $9525$ cycles, and is thus usually slower to execute than the \texttt{mov} in the \emph{smaller-than} path.
Consequently, if a measured \texttt{mov} timing is ``slow'' it must mean that the \emph{bigger-than} path was executed (3\% false positive). 
Overall, by using this comparison repeatedly with a secret bitstring as input, we were able to accurately recover ~25\% of the secret's bits (with $1000$ function calls).

\subsection{Montgomery Modular Multiplication}
\begin{figure}[t]
    \centering
    \includegraphics[trim={0.2cm 0.4cm 0.2cm 0.2cm},clip,width=0.9\linewidth]{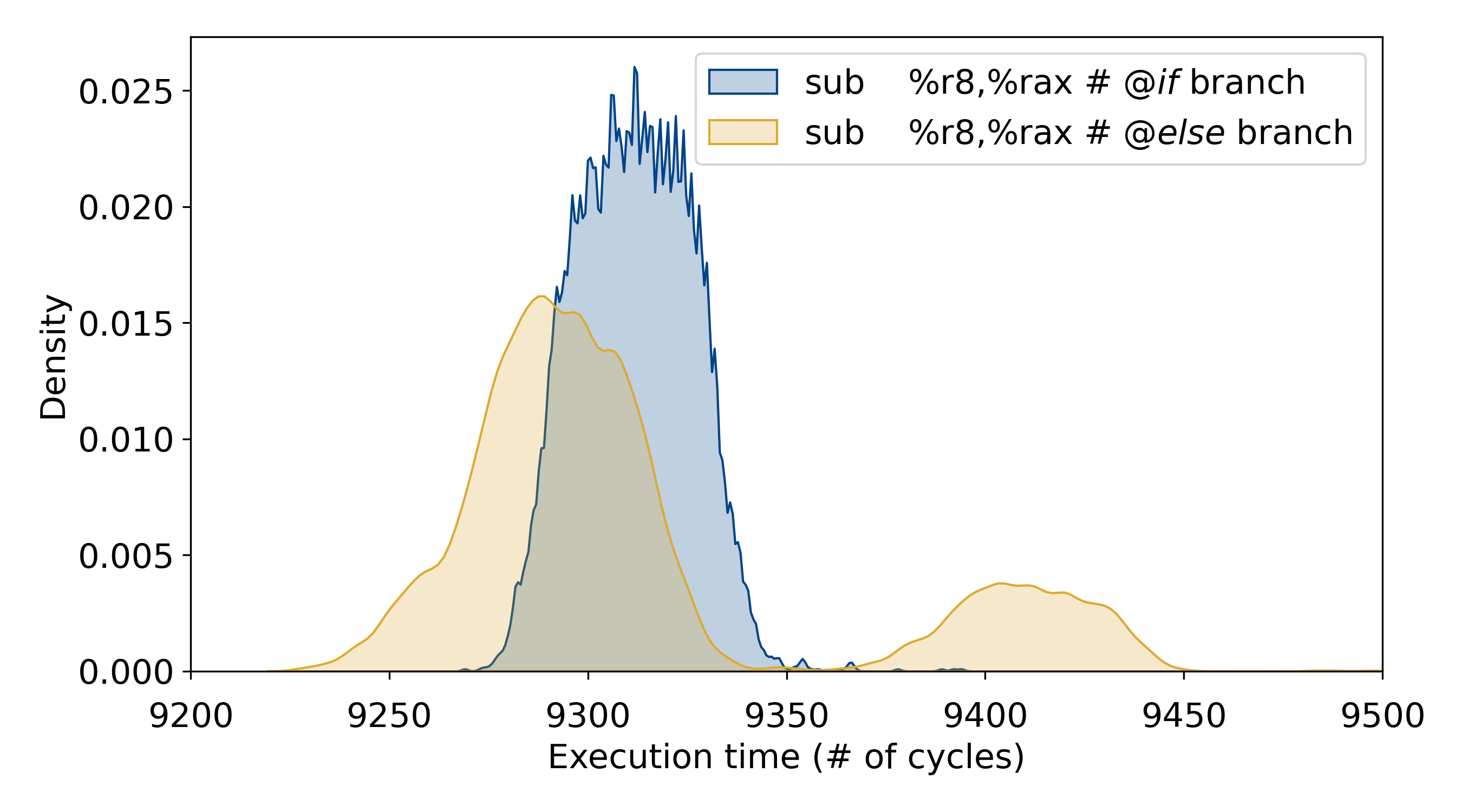}
    \caption{Comparison of the real subtraction (\emph{if} branch) and dummy subtraction (\emph{else} branch) branches in the mbedTLS MM implementation. The two branches are identical, and both include a \texttt{for loop} that executes two memory writes (cf.\ Listing~\ref{lst:mbedtlasm}). The graph shows the distribution of the 11th instruction in the \texttt{for loop}  (a \texttt{reg} to \texttt{reg} \emph{subtraction}), highlighting that as long as memory writes are present, surrounding loop instructions produce different distributions based on their alignment as well. The distributions were estimated from $1000$ function calls, each of which has $6$ loop iterations, resulting in $6000$ measurements per instruction.}
    \label{fig:mbedtlsattack}
    \vspace{-1em}
\end{figure}

The Montgomery modular multiplication (MM) is a fast MM algorithm often used in cryptographic libraries due to its efficiency and minimal secret dependence. There is only a single secret-dependent branch in the algorithm: a conditional subtraction that is done at the end of the multiplication. MM is used to perform modular exponentiation, and knowing whether the subtraction was done or not leaks some bits of a secret key used in the exponentiation~\cite{montmul_expleakage}.
Some implementations, including mbedTLS as of version 2.16.6, just balance the branches by adding an else branch with a dummy subtraction in it (cf.\ Listing~\ref{lst:mbedtlscost}). However, this naive mitigation is still vulnerable to side-channel attacks that target control-flow secret dependencies, such as the \attackname attack.
We compiled the mbedTLS library with the \texttt{gcc -O3} flag and used it inside an enclave that performs a modular exponentiation (as the MM function is not directly exposed in the library's API). The \texttt{O3} flag inlines functions when possible, so instead of performing two function calls, as shown in Listing~\ref{lst:mbedtlscost}, the binary contains two identical copies of the \texttt{mpi\_sub\_hlp} function. The branch condition determines which of these two gets executed.
The \texttt{mpi\_sub\_hlp} function contains a loop with two memory writes. The loop repeats a number of times proportional to the size of the modulus of the multiplication. In Listing~\ref{lst:mbedtlasm} in the Appendix, we give the assembly code generated by the compiler for the loop we exploit. Since the two loops were aligned differently, they exhibited different timing distributions, as shown in Figure~\ref{fig:mbedtlsattack}. While the differences were not as big as seen in our controlled tests (most likely due to the fact that several instructions are executed in between consecutive memory writes), they were enough to differentiate the branches. Using Welch's t-test, we correctly classified 83\% (511 out of 616) subtraction calls (whether they were dummies or not) with 99.9\% confidence with just 16 repetitions of an exponentiation with the same inputs. %

\subsection{Leaking RSA Keys}
We demonstrate a full end-to-end attack leveraging the \attackname attack by exploiting the  function that generates a new random RSA key pair (\texttt{mbedtls\_rsa\_gen\_key}) in mbedTLS v2.16.6. This function has several secret-dependent branches. The one we target is executed during the computation of $gcd(e, (p-1)(q-1))$, where $e$ is the RSA public exponent and $p$ and $q$ are two RSA primes. Leaking $(p-1)(q-1)$ allows to easily compute the RSA private key (as together with $n=pq$ we can solve for $p$ and $q$ and then compute $d = e^{-1} \mod{\lambda(n)}$). Control-flow leakage from the \texttt{gcd} implementation has been thoroughly studied~\cite{gcd_leakage, gcd_branch_leakage, gcd_leak} but  it only leads to partial information recovery without fine grained execution traces~\cite{gcd_leakage}. The binary \texttt{gcd} implemented in mbedTLS has a main loop that removes the trailing zero bits to its operands and then has a balanced branch in which a subtraction and a \texttt{shift--right} is performed. 
To recover the RSA private key it is sufficient to leak two pieces of information: the output of the function that counts the number of trailing zero-bits and the path taken in the balanced branch.
We leak the trailing zero-bits by counting the number of instructions executed in the respective function, as demonstrated in~\cite{moghimi2020copycat, gcd_leak}. The result of the balanced branch is leaked with the \attackname attack. Similar to the MM attack described above, the branches need to contain inlined function calls for the attack to work. To achieve this, we modified the signature of the \texttt{int mbedtls\_mpi\_shift\_r(...)} function in \texttt{bignum.c} to \texttt{\textit{inline} int mbedtls\_mpi\_shift\_r(...)}. Note that different compiler versions might lead to the compiler inlining this function on its own, thus producing a vulnerable binary.
With this function inlined, the branches both contain the loop shown in Listing~\ref{lst:mbedrsa} in the Appendix, leading to a differently aligned memory write depending on the branch taken. 
This loop within the branches is usually executed $32$ times, giving us a fairly high number of memory writes to profile. We collect and use the information from the distribution of each instruction in the loop in order to recognize which branch is being executed. The overall timing distributions are omitted here due to lack of space, but in short some instructions look like Figure~\ref{fig:mbedtlsattack}, while others more like Figure~\ref{fig:ipp_attack}. This means that we can classify the branch whenever any instruction's timing is in the ``slow'' mode of Figure~\ref{fig:ipp_attack} or whenever an instruction's timing is in the tail of the distributions of Figure~\ref{fig:mbedtlsattack}. 
We executed $1000$ runs, fixing the exponent to $e = 65537$, and generating a new pseudo-random key in each run. Note that since a new key is generated on each run, we cannot correlate the executions of multiple runs. In each execution, the attacked branch was executed $1018$ times on average ($std = 25.40$), and on average we could not classify $89$ ($std = 92.35$, median = $55$) branches. This means that on average we would need to brute force $89$ bits to recover the secret key.
In practice, we noticed that since the exponent is orders of magnitude smaller than $(p-1)(q-1)$, early iterations of the secret branch are very likely not taken. Leveraging this information, we perform several guesses of the key starting from the last unclassified iteration. We assign this iteration as `taken` and check if this results in a correct key. If not, we assign the next iteration as taken as well and repeat. This greedy approach worked on 65\% of the runs and allowed us to recover the key of those runs in a matter of seconds.

\section{Affected Processors and Configurations}\label{sec:affected_cpu}

\begin{table}[!t]
    \centering
    \resizebox{\linewidth}{!}{%
    \begin{tabular}{@{}lllllc@{}} \toprule
          Processor    & $\mu$arch       & Launched & $\mu$code & Mitig.    & Vulnerable \\ \midrule
          i7-6700HQ    & Skylake         & Q3'15    & 0xc2      & $\mu$code & yes$^\dagger$ \\
          i7-6700HQ    & Skylake         & Q3'15    & 0xd6      & $\mu$code & yes$^\dagger$ \\
          i7-7700      & Kaby Lake       & Q1'17    & 0x48      & -         & yes \\
          i7-7700      & Kaby Lake       & Q1'17    & 0x8e      & $\mu$code & yes$^\dagger$ \\
          i7-9700K     & Coffee Lake R   & Q4'18    & 0xb8      & HW        & yes \\
          i7-9700K     & Coffee Lake R   & Q4'18    & 0xca      & HW        & yes \\
          i9-9900KS    & Coffee Lake R   & Q4'19    & 0xb8      & HW        & yes \\
          i9-9900KS    & Coffee Lake R   & Q4'19    & 0xca      & HW        & yes \\
          i9-10900K    & Comet Lake      & Q2'20    & 0xca      & HW        & yes \\ 
          Xeon E-2278G & Coffee Lake R   & Q2'19    & 0xb8      & HW        & yes \\
          Xeon E-2278G & Coffee Lake R   & Q2'19    & 0xca      & HW        & yes \\
         \bottomrule     
    \end{tabular}}%
    \vspace{2px}
    {\raggedright\footnotesize\quad\enskip\,$^\dagger$ Only vulnerable in some runs (see Figure~\ref{fig:succcess_rate_old_cpu})\par}
    \vspace{-1em}
    \caption[]{List of all the processors we tested with their respective microcode version. The \emph{Mitig.} column indicates whether the mitigation against known microarchitecural attacks such as Spectre and Foreshadow is implemented in hardware (HW) or $\mu$code.}\label{tab:microcode}
    \vspace{-1em}
\end{table}

We tested five different processors from the 6th generation, which introduced Intel SGX, up to the 10th which has hardware mitigations for recent microarchitectural attacks~\cite{hardwaremitigations}. We give the details of the CPUs tested in Table~\ref{tab:microcode}.
For each processor, we tested the minimum microcode version supplied by the mainboard and the most up to date version as of February 2020. Each CPU was tested by computing the attack success rate for various alignments as in Figure~\ref{fig:heatmap}.
The \attackname attack was successful on all tested CPUs and microcodes. %

\begin{figure}
    \centering
    \includegraphics[trim={8 8 8 6},clip,width=\linewidth]{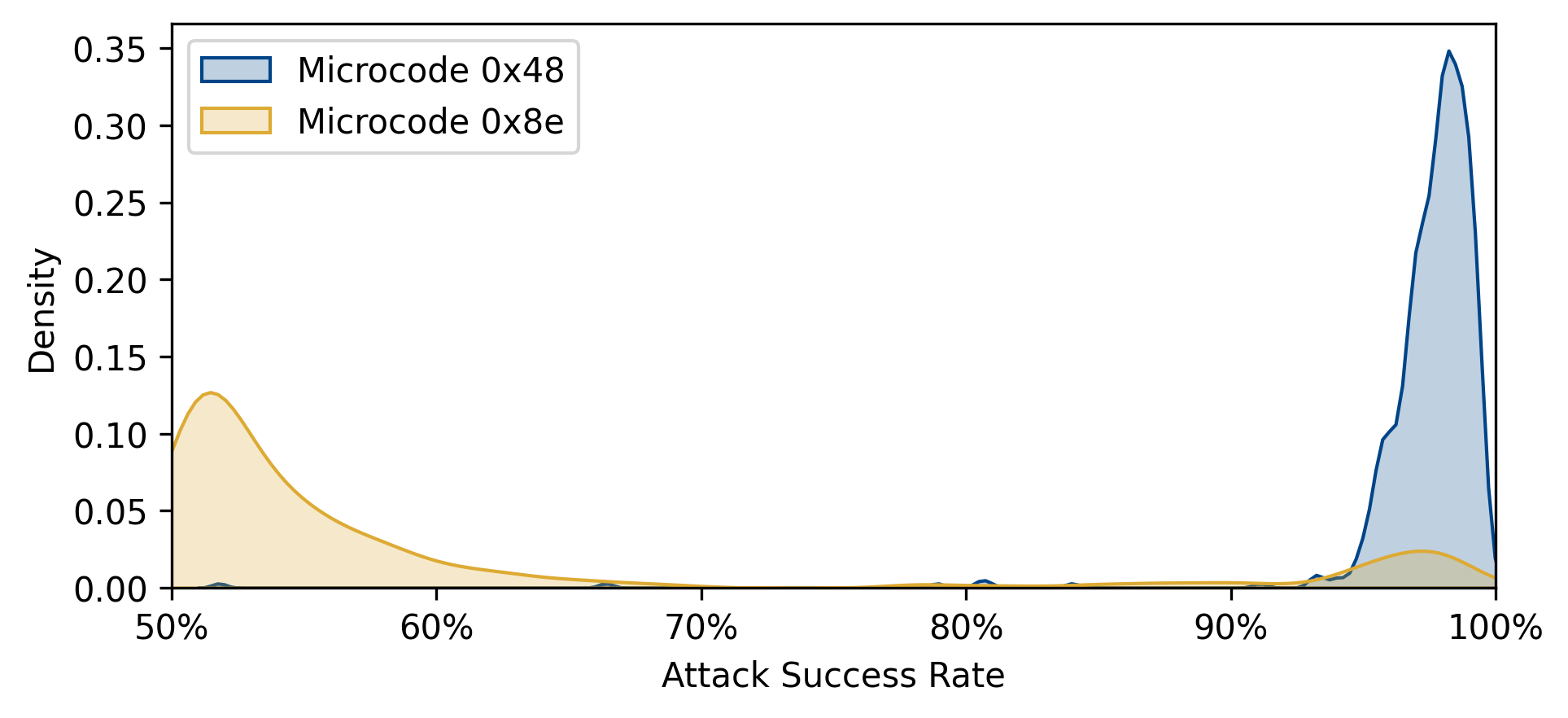}
    \vspace{-10pt}
    \caption{Distribution of the attack success rate with different microcode versions of an Intel Core i7-7700 CPU - across $500$ runs per microcode. For each run, we estimate the attack success rate as the percentage of branches the attacker guessed correctly among $1000$ executed branches from Figure~\ref{fig:asm_long_code}, with alignment $X=6, Y=2$. }
    \label{fig:succcess_rate_old_cpu}
    \vspace{-1.0em}
\end{figure}

Our measurements indicate that the processors can be separated into two groups with similar behavior: processors with and without hardware mitigations against various microarchitectural attacks. Interestingly, newer processors with hardware mitigations built-in were more susceptible to our attack, whereas older processors with mitigations in microcode seem to add noise and thus have lower success rates on average.
More in-depth analysis revealed that the most recent microcodes on processors without hardware mitigations increase the number of cycles used for \texttt{AEX} and \texttt{ERESUME} and add some randomness to our experiments. For these configurations, every run of the experiment exhibits a different behavior. Figure~\ref{fig:succcess_rate_old_cpu} shows the success rate for $500$ separate runs each with $1000$ samples. Note that most of the runs with the new microcode show a random success rate. However, some runs exhibit a clear timing difference leading to a $>95\%$ success rate. The adversary can detect which behavior a particular run is going to exhibit by observing the timings of early \texttt{mov}s aligned at particular addresses. Thus they could decide whether to attack or not before the secret is retrieved or provisioned, and relaunch the enclave until its behavior is clearly vulnerable.

\section{Potential Causes}\label{sec:causes}
The complexity of the microarchitecture of current Intel processors makes it very challenging to pin-point the cause of the timing differences to a specific component. However, we will discuss some components which we were able to decisively exclude. We start with the memory subsystem, then we investigate the execution engines, and finally we will focus on the frontend. For each potential culprit in these building blocks, we will describe an initial theory and then try to refute or confirm it using performance counters and other measurements. Note that the performance counters are sparsely distributed over the entire core and do not exhaustively cover the entire microarchitecture. Therefore, investigation into some hypotheses is very challenging if no performance counters exist for the respective part of the processor.

\paragraph{Memory Subsystem}
Observation~\ref{obs:write_fast_slow} and \ref{obs:memory_writes} point to potential causes in the memory subsystem. Specifically the fact that the slow \texttt{mov} is around 100 cycles slower. For a current-generation processor, 100 cycles is a rather large delay that is usually only observed for accesses to external memory or the last level cache. %
However, performance counters refute any theory related to the memory subsystem since all performance counters related to external memory or last level cache did not show a difference between the slow and the fast \texttt{mov}s.

\paragraph{Execution Engines}
The execution engine gets a list of instructions from the allocation queue as input and tries to reorder and execute them as fast as possible. As far as we know, it is completely decoupled from the frontend and does not depend on any alignment since it works on decoded micro-ops. However, given Observation~\ref{obs:mod16}, we know that the alignment influences the timing difference. We thus rule out the execution engine as the root cause of the timing differences.

\paragraph{Frontend}
Observation~\ref{obs:mod16} strongly hints at the frontend as the culprit, since the fetch window is one of the only structures which operates at a 16 Bytes granularity, matching the 16 Bytes periodicity of the observations. %
 
The micro-op cache is a microarchitectural structure in the frontend~\cite{solomon2003micro} that holds previously decoded fetch windows and serves them to, for example, repeated jumps to the same address. 
On a micro-op cache hit, many cycles can be saved due to not having to decode the instructions again. Our observed timing difference might stem from hits and misses in this cache. For some interrupts, the micro-op cache might miss, and the instructions must be decoded again. Whereas, for some others, it hits and immediately proceeds to the reorder buffer. However, the timing difference we observed seems excessively large for this kind of small difference in the execution path. Besides, performance counters that measure the behavior of the micro-op cache show an equivalent number of hits in the slow and the fast \texttt{mov}s. Thus, we rule out the micro-op cache as a cause.

Branch prediction is responsible for predicting the future control flow. The core will fetch ahead and speculatively continue to execute in the predicted path. Branches and jumps where the target is not immediately known (e.g., the target comes from memory) both rely on the branch predictor to guess which instruction will be executed next. Hence, the resumption of the enclave could potentially suffer from a misprediction on the current enclave instruction and therefore suffer from a delay. However, all performance counters that we measured did not show any additional mispredictions for slow or fast instructions. 

\paragraph{Summary} While we were able to decisively refute many of the most common reasons for timing differences, none of our tests were able to identify with reasonable confidence an explanation for the observed timings exploited by the attack.  %

\section{Defenses}

There exist various defenses against the \attackname attack, some of which we will discuss in this section. First and foremost, we want to stress that data-oblivious code~\cite{rane2015raccoon,practicalmiticoppens09} is a principled approach that thwarts every known side or controlled-channel attack. We discuss these techniques in Appendix~\ref{sec:data-obliv}. As such it also remains secure against the \attackname attack. Nevertheless, data-oblivious code presents several challenges in practice, as it is hard to get right and results in high overhead in certain applications. Therefore, in practice, many spot defenses against the known attacks have been used since they are usually easier to apply and more performant. However, most of these spot defenses are circumvented by new attacks such as the \attackname attack. 
While the behavior exploited by the \attackname attack stems from the underlying hardware, the simple defense we discuss is at the software level. Hardware mitigations would also be possible, but due to the lengthy turn-around time for new processors, software defenses are more attractive. %

As seen in Section~\ref{sec:obs}, the execution time of individual instructions depends on their alignment. %
Particularly, branches with identical alignment do not exhibit any observable timing difference. Therefore, aligning the two branches to the same address (modulo 16) leads to indistinguishable timing distributions for both branches. We evaluated the overhead in terms of binary size and performance of this approach on three common libraries: libc, OpenSSL, and mbedTLS. We used GCC v7.5.0 with the compile flag \texttt{-falign-jumps=16} - this flag aligns all branch targets to $0x10$, thwarting our attack. The highest size overhead (3.73\%) was on one of the binaries generated for libc, this however was the only outlier as all the other binaries had an overhead of less than $0.5\%$. For comparison, compiling with \texttt{-03} added on average $14\%$ compared to \texttt{-02}. To evaluate performance, we use libc-bench\footnote{https://www.etalabs.net/libc-bench.html} for libc and the benchmarks that come with the libraries for mbedTLS and OpenSSL. The \texttt{strstr} test in libc-bench had the highest overhead at 30\%, and libc overall had an average overhead of 1\%. Depending on the evaluated cryptographic function mbedTLS had overheads ranging from 4\% to -5.5\%, while OpenSSL from 3\% to -4\%, showing that for some cryptographic functions' implementation the defense even provides performance boosts.

\section{Related Work}\label{sec:rel_work}
We compare our attack and related ones in Table~\ref{tab:rel_work}. In short, the main differences lie in the type of branches that are vulnerable to the various attacks. Previous defenses build either on the fact that controlled channel attacks cannot leak at sub-page granularity or that BPU attacks cannot leak the target virtual address of unconditional branches. In general, these defenses are ineffective against our attack since we exploit a fundamentally different mechanism. In the following, we describe the differences between the \attackname attack and other related attacks in more detail.

\begin{table*}[ht!]
        \centering
        \resizebox{\linewidth}{!}{%
        \begin{tabular}{@{}llllllll@{}} \toprule
            Attack type / Name                                                                                  & Data        & CF              & Resolution                  & Synchronization with Victim   & Vulnerable branches                         \\ \midrule
            Cache~\cite{highrescacheattacksgx,kari2017softwaregrandexposure,l1sgxcacheattack,cacheattacksgx17}  & \Yes        & \Yes             & 64 B (CL)                   & Interrupt / SMT / Multicore   & If paths in different CLs                   \\
            BPU~\cite{lee2017inferring,evtyushkin2018branchscope,bluethunder}                                   & \No          & \Yes             & Branch                      & Interrupt / SMT               & If paths virtual addresses are known                   \\
            TLB~\cite{tlbattack,leakycaulderon17}                                                               & \Yes         & \No              & 4 KiB (Page)                & SMT                           & If different data pages accessed based on path        \\
            False Dependency~\cite{cachebleed,memjam}                                                           & \Yes         & \No              & 4 B                         & SMT                           & If data $> 4B$ apart is accessed based on path \\
            Port contention~\cite{portcontentionsgx,microscope}                                                 & \No          & \Yes             & $\mu$ops                    & SMT                           & If paths issue different $\mu$ops           \\
            PT Controlled-Channel~\cite{ccpagefault,accesscontrolledchannel,leakycaulderon17,offlimits18}       & \Yes         & \Yes             & 4 KiB (Page)                & Page-Fault / Interrupt / SMT  & If paths in different pages                 \\
            Nemesis~\cite{van2018nemesis}                                                                       & \No*        & \Yes             & Instruction type and count  & Interrupt                     & If paths have different instructions        \\
            CopyCat~\cite{moghimi2020copycat}                                                                   & \No          & \Yes             & Instruction count           & Interrupt                     & If paths have a different instruction count \\
            \textbf{\attackname attack}                                                                             & \textbf{\No} & \textbf{\Yes}    & \textbf{Instruction VA}     & \textbf{Interrupt}            & \textbf{Any branch (Must have a store)}     \\
            \bottomrule  
        \end{tabular}}%
        \vspace{2px}
        {\raggedright\footnotesize\quad\,* Leaks instruction operands (if they induce different execution time). E.g., multiplication to $1$ vs. multiplication with big numbers.\par}
        \vspace{-1em}%
        \caption[]{Overview and comparison of related SGX side-channel attacks. The first two columns indicate whether the attack can leak data-dependent or control-flow (CF) dependent secrets. %
        The \attackname attack is the only attack that can leak the decision made for any type of branch (as long as they contain a memory store in them), even if they are based on indirect unconditional jumps (e.g., as a mitigation against BPU attacks), or if both paths are contained within the same CL (e.g., as a mitigation against cache and controlled-channel attacks).
        }\label{tab:rel_work}
        \vspace{-0.5em}
\end{table*}

\subsection{Controlled-Channel Attacks}

The attacker's control over the OS enables novel noise-free deterministic side-channels~\cite{ccpagefault,leakycaulderon17,accesscontrolledchannel} known as controlled-channels since the attacker controls the channel. Memory paging, the scheduler, the handling of interrupts and exceptions, are a few examples of what the attacker can take advantage of -- every interface between the OS and the enclaves can be leveraged in controlled channel attacks.
In~\cite{ccpagefault}, Xu et al.\ modify page permissions so that the CPU generates a page fault for each page the enclave tries to access. The trace of page faults contains enough information to, e.g., let attackers reconstruct images processed in the enclave. Subsequent attacks made controlled channel attacks stealthier, by observing that the CPU sets the accessed and dirty bits~\cite{accesscontrolledchannel,leakycaulderon17} in the page tables (PTs), thus allowing to monitor the enclave's execution without having to trigger page faults.
However, the resolution of page-based controlled channel attacks is quite coarse, allowing the attacker to know only whether any access in a page (4 kB) was made, but not where within it.

The coarseness of PT based controlled channel attacks is an element that defenses have latched onto, to protect enclaves~\cite{shinde2016preventing,sgxshield}. These defenses either call for sensitive code to be within a page~\cite{shinde2016preventing} or randomize the enclave's page layout so that page accesses cannot be correlated~\cite{sgxshield}. Even Intel specifies that controlled channels can be mitigated ``by aligning specific code and data blocks to exist entirely within a single page''~\cite{sgxsidechanneldev}.
However, the resolution of controlled channel attacks was increased through an attack exploiting legacy memory segmentation~\cite{offlimits18}, which is also managed by the OS. While the attack only works under uncommon circumstances (32 bit enclaves and smaller than 1 MiB), it can observe memory accesses at 1 byte granularity.

Our attack can trace the control-flow of an enclave with instruction granularity, thus increasing the resolution of PT-based controlled channel attacks. %
Like other controlled channel attacks~\cite{van2018nemesis,moghimi2020copycat}, the \attackname attack relies on interrupts to observe instructions and control-flow within a page. However, it differs from them on the kind of branches that it can exploit. Nemesis~\cite{van2018nemesis} can distinguish between branches that have instructions with measurable timing differences, either because they have different kinds of instructions in their paths, or because they have a different number of instructions. CopyCat~\cite{moghimi2020copycat} can track the control-flow in branches with a different number of instructions. The \attackname attack allows differentiating any branch, even if both paths contain the very same instructions and are hence not vulnerable to other controlled channel attacks. The only requirement for our attack is that the branch contains at least a memory store in it. Such higher resolution hence defeats previous defenses that rely on controlled channels being limited to observe only at a page resolution.

\subsection{Microarchitectural Side-channel Attacks}
Microarchitectural attacks exploit information leakage due to shared microarchitectural resources across different privilege domains. Among these shared resources, the ones that have been exploited the most are the cache and the branch prediction unit (BPU). We examine side-channel attacks based on these and other shared microarchitectural components below.

\paragraph{BPU Attacks}
The BPU records the outcome of recent branches and jumps, to aid the CPU speculation. As it is shared among different execution contexts running in the same core, it can leak information about the control-flow of another context. The BPU was the focus of recent attacks, and particularly against SGX~\cite{lee2017inferring,evtyushkin2018branchscope,bluethunder}. 
BPU attacks require either SMT~\cite{bluethunder} or time multiplexing at a fine granularity between the victim and the attacker in the same physical CPU core~\cite{bluethunder,evtyushkin2018branchscope,lee2017inferring}. These attacks are, in general, very sophisticated, and require reverse-engineering the BPU. Given how hard this is to achieve, BPU attacks are not easy to generalize to different microarchitectures and to pull off in practice~\cite{deadkaslr}. These attacks are also limited to the type of branches they can exploit. For instance, they cannot leak the target virtual address of indirect jumps~\cite{lee2017inferring}.
As these attacks give fine-grained information to the attacker, there have been a few defenses proposed against them~\cite{lee2017inferring,bluethunder,branchshadowmitig}. Most notably, some defenses call for a holistic approach by flushing the BPU accross context switches~\cite{lee2017inferring,bluethunder}. Other defenses propose spot defenses such as replacing every branch with indirect jumps~\cite{branchshadowmitig}.
BPU attacks are particularly related to the \attackname attack, as they both exploit secret-dependent branches. However, as the \attackname attack exploits a fundamentally different mechanism, any spot-defense against BPU attacks is not effective against our attack.

\paragraph{Attacks on caches and other shared resources}
Because caches are a resource shared across different execution contexts, an attacker thread can infer which accesses a victim recently made in another context by obtaining information about the cache state.
While cache attacks often exploit timing variations in access latency to probe the state of the cache~\cite{timingsidesurvey}, state changes can also be detected by using instructions' side effects~\cite{primeabort, cachenotime}. Cache attacks target different levels of the cache hierarchy -- from core-local data cache~\cite{aciiccmez2007yet,osvik2006cache,aciiccmez2008vulnerability,l1cacheattack,l1datattackaes,kari2017softwaregrandexposure,l1sgxcacheattack,highrescacheattacksgx,cacheattacksgx17} and core-local instruction cache~\cite{aciiccmez2007yet,l1cacheattack}, to the last level cache (LLC) which is shared amongst all cores~\cite{yarom2014flush,gruss2016flush,sgxcacheattack}. As code and data are shared in the upper levels of cache (from L2), attacks that exploit them can leak both control-flow-dependent and data-dependent secrets~\cite{yarom2014flush,gruss2016flush,sgxcacheattack}.
Attacks on core-local caches require to be co-located with the victim and thus usually rely on simultaneous multithreading (SMT) or on accurate time-multiplexing. On the other hand, attacks that exploit the LLC can be run at the same time as the victim in another core.

The TLB is a shared buffer that stores the translation information from virtual addresses (VA) to physical addresses. It can be exploited to detect whether a victim recently accessed a data memory page~\cite{tlbattack,leakycaulderon17}. Since the TLB is shared only among processes in the same core, it has been exploited only using SMT so far. It can leak data accesses at a 4 kB granularity.
CacheBleed~\cite{cachebleed} was the first attack to demonstrate intra-CL leakage for data accesses, achieving a resolution of 8B. It exploited 
cache bank conflicts and write-after-read false dependencies. Since the adversary is not in the same address space, they induce a false memory dependency by making use of 4k page aliasing - where an address $x$ is considered the same as $x+4096$ by the hazard detection in the processor. Cache banks are only present in older Intel architectures and therefore cannot be exploited on newer CPUs.
Moghimi et al.~\cite{memjam} ported the CacheBleed attack to newer CPU and SGX while improving the resolution to 4B in their MemJam attack. They exploit read-after-write false dependencies in the processor memory subsystem using 4k aliasing.
The PortSmash~\cite{portcontentionsgx} attack extended the resolution available to the attacker even further, by being able to detect issued microops in SGX enclaves. It works by keeping specific CPU execution ports busy and monitoring their execution latency. Execution in these ports becomes slower when another context is using them, thus leaking information about their control-flow to the attacker.

\section{Conclusions}
In this work, we observed a dependency between instructions execution time and their alignment modulo 16. We attributed these differences to the CPU frontend and its fetch and pre-decode module. We leveraged these time dependencies to construct the \emph{\attackname} attack, which can leak the instruction pointer of an SGX enclave at the byte level granularity. The \attackname attack works against any kind of branch, as long as it contains at least a memory write. It can attack perfectly balanced branches, even when they fit within one cacheline. We showed that the \attackname attack achieves a success rate of more than 99\%, depending on the target victim code.
We tested every modern CPU microarchitecture that currently supports SGX (up to 10th gen) and found them all to be vulnerable to our attack. We demonstrated the practicality of our attack by exploiting two commonly used cryptographic libraries, mbedTLS, and the Intel IPP Cryptography library. We discussed relevant defenses to the attack, such as aligning all branch targets to the same offset modulo 16. While we show that this defense has tiny size and performance overheads, we stress that, in general, secret-depending branching should be avoided to guarantee confidentially in SGX enclaves.

\ifsubmission
\else
\section*{Availability}
A proof of concept of the attack is available online at \url{https://github.com/dn0sar/frontal_poc}.

\section*{Acknowledgements}
We would like to thank Kaveh Razavi for insightful discussions about the root causes of the \attackname attack and Kari Kostiainen for his feedback on early drafts of this paper. We thank our shepherd Yuval Yarom and the anonymous USENIX reviewers for their valuable suggestions.
\fi

\renewcommand*{\bibfont}{\small} %
\printbibliography

\appendix
\section{Supplemental information}

\subsection{Responsible disclosure}\label{sec:disclosure}
We notified the Intel PSIRT on February 21 2020, about the \attackname attack. We sent them a previous version of this paper and a proof of concept for the vulnerabilities we identified.
They informed us on April 22 that their best practices~\cite{sidechannelbestpractices} already suggest avoiding secret-dependent branching, and therefore our attack is considered out-of-scope for their SGX libraries. In particular, they stated that the balanced branches of the IPP Crypto library we attack in Section~\ref{sec:ipp_intel_vuln} are not used for secret-dependent operations in the SGX architectural enclaves and hence do not pose any security implication.
The vulnerability shown in Listing~\ref{lst:mbedtlscost} was reported to the mbedTLS team%
, which promptly fixed it. The vulnerability was also described in a 2017 paper~\cite{lee2017inferring} and was still unknown to the developers.

\subsection{Data-oblivious Execution}\label{sec:data-obliv}
Resilience against side-channel attacks is often a desired security property when implementing software. This property is particularly important for libraries and applications that operate on secret and sensitive data on a system controlled by the attacker. 
Side-channel attacks exploit secret-dependent variations of the program execution. These variations are generally of two types: control-flow dependent and data-dependent.
Control-flow secret dependencies are present whenever the control flow of an application depends on confidential information.
Data dependencies manifest when latency or resources utilized depend on the input data. For example, when memory accesses at different addresses are performed based on some secret.
Countless attacks have exploited these types of dependencies in the past~\cite{kocher1996timing,percival2005primeprobe,yarom2014flush}, targeting in particular cryptographic libraries, as extracting secret keys handled by these libraries breaks any security guarantee built on top of them.
Data oblivious execution defends against side-channel attacks by removing the two dependencies mentioned above. This eliminates any variation in program execution that would be potentially observable by the attacker.
There are two ways to obtain a data oblivious executable - first writing it directly in low level assembly code, second by performing an automatic transformation at compile time from a higher level language. Note that writing the code in a higher level language in a data oblivious way, and then simply compiling it, might reintroduce data or control flow dependencies at the binary level.

Several techniques for compiling and transforming code from an arbitrary high level language to data oblivious code have been proposed~\cite{rane2015raccoon, liu2015ghostrider, molnar2005program, liu2013memory, practicalmiticoppens09}.
One of the most complete constant-time transformation for SGX is Raccoon~\cite{rane2015raccoon}. %
It removes any control flow and most data dependencies by transforming secret-dependent branches into a decoy and a real path that contain similar instructions. At run time, both paths are executed, allowing only the real one to modify memory by carefully applying the conditional move instruction (\texttt{cmov}). Raccoon runs on SGX enclaves and uses SGX's memory protections to ensure confidentiality against an attacker that can otherwise read arbitrary locations of memory.

\subsection{Measurement Details}\label{sec:setup}

\begin{figure}
    \centering
    \includegraphics[width=0.9\linewidth]{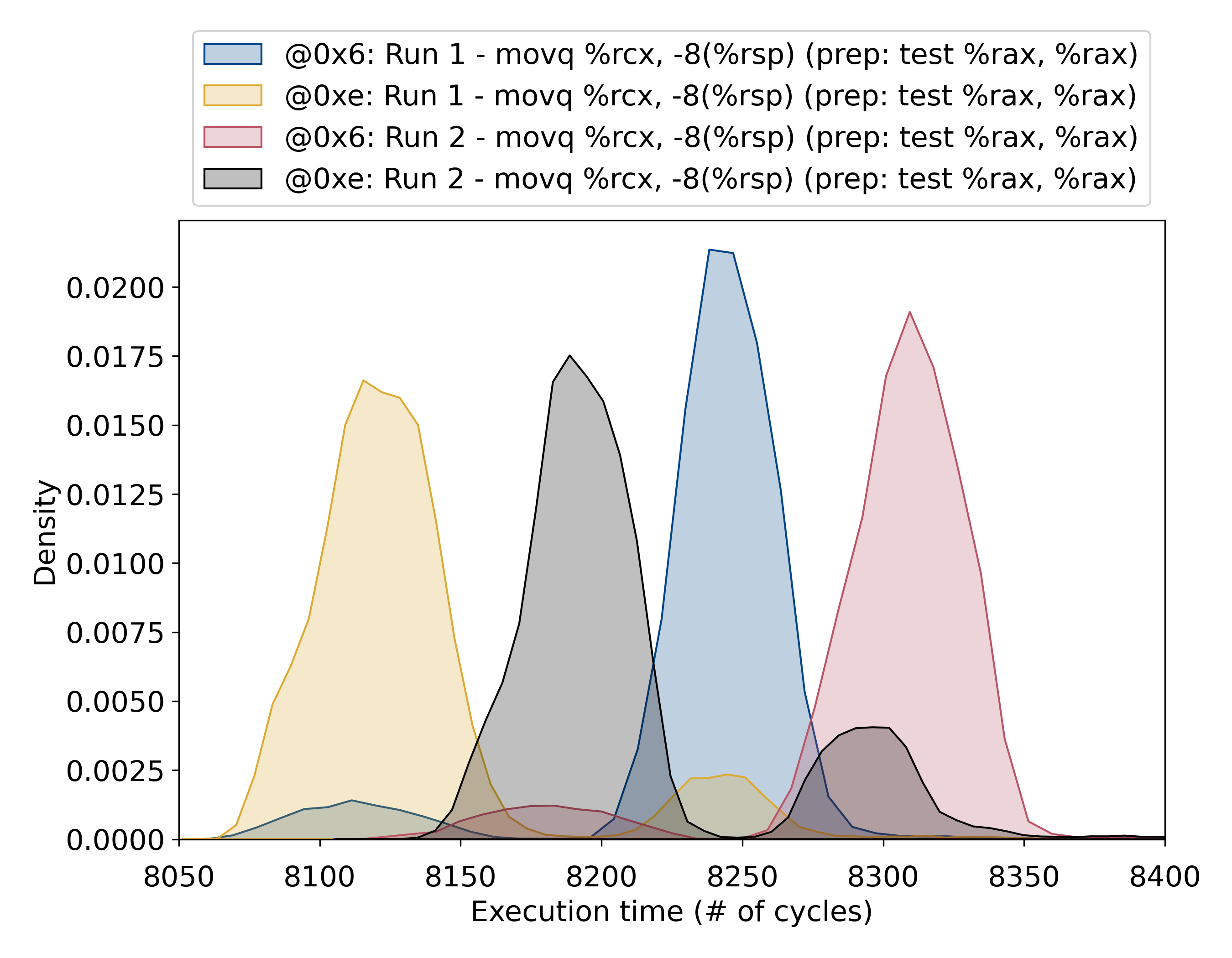}
    \vspace{-10pt}
    \caption{Distribution of the instructions of Figure~\ref{fig:long_seq_split} across different runs, split by their alignments. This figure highlights how different enclave runs exhibit a shift in the mean of their instructions' distribution and hence distributions are not directly comparable between runs.}
    \label{fig:different_runs}
\end{figure}

We made several changes from stock SGX-Step~\cite{van2017sgx}, primarily aiming to reduce measurement noise as much as possible. In terms of functionality, we added the possibility to measure performance counters alongside instructions' timings.
We identified four major sources of noise: the OS, variability in the APIC timer, unpredictability of shared resource state, and enclave creation offset noise. We discuss how we addressed each of these in the following paragraphs. 

The OS is a source of noise as it needs to run the scheduler on each core to decide which tasks to execute. If the scheduler runs in between the start of a measurement and its end, the measurement will inevitably be longer. Moreover, running any OS function while we single-step can sometimes evict part of the enclave memory from the cache, thus forcing the enclave to fetch it again when it is resumed. This also happens when the scheduler executes on the sibling core. As recommended in the original SGX-Step framework, we run the code in its own isolated core to reduce this noise. However, this alone stops neither the scheduler nor the other cores from interrupting the isolated core. We observed that disabling watchdogs at boot and disabling the graphical user interface tends to reduce the noise produced by the OS, albeit it does not eliminate it completely.

In stock SGX-Step, the APIC timer is set in the \texttt{aep\_cb\_func}. The \texttt{aep\_trampoline} then executes and resumes the enclave. Various conditions can create variability between the time in which the APIC timer is set and the time at which the enclave resumes. For instance, sometimes, the \texttt{aep\_trampoline} code page or some of the data it uses might not be present in the cache. We addressed this variability by setting the APIC timer from the \texttt{aep\_trampoline} function with a value passed from the \texttt{aep\_cb\_func} function and by serializing the instruction stream (using \texttt{CPUID}) just before setting the timer. Interestingly, while debugging for this source of noise, we observed that we were never able to interrupt in between fused macro-instructions as these seem to be treated atomically by the CPU, as also observed in~\cite{moghimi2020copycat}.

The third source of noise stems from the difference in the microarchitectural state in-between measurements. While we could not completely eliminate this source of noise as we have no direct view of the microarchitecture, we were able to reduce it significantly. The most effective change in this regard was obtained by linearizing the code of the \texttt{aep\_cb\_func} so that there is no mis-speculation in between single-steps and the function always has the same cache footprint. Even with this change, we observed that instructions that cross a virtual page boundary remained noisy. To account for this, we remove these measurements from the trace when possible. Note that the attacker can easily tell if an instruction crosses the page boundary as the access bit of the new page is set by the CPU.

Finally, while validating these changes, we noticed a source of noise across enclave creations, whose effects we illustrate in Figure~\ref{fig:different_runs}. The figure shows the measurement of the \texttt{mov}s from Figure~\ref{fig:long_seq_split} across enclave creations. As can be seen, the distributions remain bimodal, but the position of the modes across creations changes. However, the relative position between the modes stays the same: the \texttt{mov} at alignment $0x6$ is slower than that at $0xe$ on both runs. While we never observed modes shifting more than $200$ cycles, this shift is still large enough such that, for instance, the distribution of \texttt{nop} instructions could overlap with the distribution of multiplication instructions from different runs.
Given this shifting between enclave creations, we concluded that instructions' timings \emph{are only comparable within the same enclave}. %

\vspace{8pt}

\lstset{basicstyle=\ttfamily\small,breaklines=true}
\begin{lstlisting}[language={[x86masm]Assembler},caption={Exploited \texttt{for} loop in the mbedTLS library's \texttt{mpi\_montmul} function (compiled on \texttt{gcc 7.5.0} with \texttt{-O3}).},label={lst:mbedtlasm},captionpos=b,xleftmargin=0em]
loop_start:
    mov    (%rcx,%rdx,8),%rax
    xor    %r9d,%r9d
    cmp    %rsi,%rax
    setb   %r9b
    sub    %rsi,%rax
    mov    %rax,(%rcx,%rdx,8)
    mov    (%rdi,%rdx,8),%r8
    mov    %r9,%rsi
    cmp    %r8,%rax
    adc    $0x0,%rsi
    sub    %r8,%rax
    mov    %rax,(%rcx,%rdx,8)
    add    $0x1,%rdx
    cmp    %rdx,%rbp
    jne    loop_start
\end{lstlisting}

\vspace{8pt}

\lstset{basicstyle=\ttfamily\small,breaklines=true}
\begin{lstlisting}[language={[x86masm]Assembler},caption={Exploited \texttt{for} loop in the \texttt{mbedtls\_rsa\_gen\_key} function of the mbedTLS library.},label={lst:mbedrsa},captionpos=b,xleftmargin=0em]
loop_start:
    mov   (%rax), %rcx
    sub   $0x8, %rax
    mov   %rcx, %rdx
    shl   $0x3f, %rcx
    shr   $0x1, %rdx
    or    %rdi, %rdx
    mov   %rcx, %rdi
    mov   %rdx, (%rax+8)
    cmp   %rax, %rsi
    jnz   loop_start
\end{lstlisting}

\ifpublic
\else
\subsection{Outside Intel SGX}\label{sub:exceptions}
A question remains on whether these effects manifest only while executing code inside an SGX enclave or whether they are present also while running a program outside of SGX. Since we cannot send interrupts fast enough during a normal execution, we decided to simulate the effect of the interrupts by modifying the code in Figure~\ref{fig:asm_long_code} such that each \texttt{mov} triggers an exception. We handle the exception and measure the time it took to execute it, and then resume the program execution from the instruction after the one that triggered the exception.
Note that exceptions are handled very similarly to interrupts, with the key difference that the instruction that is currently executing can retire when an interrupt is triggered, while it needs to be discarded when an exception is raised.

We observed that timing differences between instructions in the two branches were less pronounced than when the code was run within SGX. Nonetheless, we were able to observe a (small) correlation between the exception handling time of single instructions and the branch being executed. This correlation hints that the effects are not only present when interrupting enclaves, but would manifest also when interrupting applications if we had a fast enough interrupt timer.
\fi

\end{document}